\documentclass[journal]{IEEEtran}

\ifCLASSINFOpdf
\else
\fi

\usepackage{graphicx}
\usepackage{xcolor}
\usepackage{amssymb,amsmath,amsfonts}
\usepackage{tikz}
\usepackage{multirow}
\usepackage{booktabs}

\usepackage{blindtext}

\usepackage{multicol}
\usepackage{tikzscale}
\usepackage{numprint}
\npdecimalsign{.}
\nprounddigits{3}

\usepackage{parskip}

\usepackage{enumitem}

\usepackage{float}

\usepackage{pict2e}

\usepackage{siunitx}
\usepackage[english]{babel}

\newcommand{\R}{\mathbb R}

\newcommand{\Ed}{\mathbf E}

\newcommand{\Hd}{\mathbf{H}}

\newcommand{\herm}{{\scriptstyle \boldsymbol{\mathsf{H}}}}
 \newcommand{\trans}{{\scriptstyle \boldsymbol{\mathsf{T}}}}

\newcommand{\ZZ}{\mathbf{Z}}                       
                       
\newcommand{\xx}{\mathbf{x}}                       
\newcommand{\zz}{\mathbf{z}}

\newcommand{\cc}{\mathbf{c}}    
                         
\newcommand{\net}{\boldsymbol{\Phi}}                         

\newcommand{\Bd}{\mathbf{B}}
\newcommand{\Ad}{\mathbf{A}}

\newcommand{\Au}{\mathbf{A}_I}             

\newcommand{\N}{\mathbb{N}} 
\newcommand{\C}{\mathbb{C}} 

\newcommand{\fun}{\mathcal{L}}             
\newcommand{\reg}{\mathcal{R}}             
\newcommand{\pen}{\Omega}             

\newcommand\norm[1]{\left\lVert#1\right\rVert}
\newcommand\set[1]{\left\{#1\right\}}

 \newcommand{\Su}{\mathbf{S}_I}                       
                          
\newcommand{\xu}{\mathbf{x}_I}                         
\newcommand{\yu}{\mathbf{y}_I}

\newcommand{\la}{\lambda}

\usepackage[percent]{overpic}
\usepackage{algorithm,algcompatible,amsmath}

\DeclareMathOperator*{\argmin}{arg\,min}
\algnewcommand\INPUT{\item[\textbf{Input:}]}%
\algnewcommand\PARAMETER{\item[\textbf{Parameters:}]}%
\algnewcommand\OUTPUT{\item[\textbf{Output:}]}%

\colorlet{lred}{red!80}
\colorlet{cmix}{blue!80!red} 
\colorlet{lgreen}{green!80}
\colorlet{lblue}{blue!80}

\newtheorem{theorem}{Theorem}
\newtheorem{definition}[theorem]{Definiton}

\newtheorem{assumption}[theorem]{Assumption}

\numberwithin{theorem}{section}

\begin{document}

\title{Unsupervised  Adaptive Neural Network Regularization for Accelerated  Radial Cine MRI}

\author{Andreas~Kofler,
        Marc~Dewey,
        Tobias~Schaeffter,
        Christoph~Kolbitsch
        and~Markus~Haltmeier
\thanks{A. Kofler is with the Department of Radiology, Charit\'{e} - Universit\"{a}tsmedizin Berlin, Berlin,
Germany (e-mail: andreas.kofler@charite.de)}
\thanks{M. Dewey is with the  Department of Radiology, Charit\'{e} - Universit\"{a}tsmedizin Berlin, Berlin,
Germany and the Berlin Institute of Health, Berlin, Germany (e-mail: marc.dewey@charite.de)}
\thanks{T. Schaeffter is with the Physikalisch-Technische Bundesanstalt (PTB), Braunschweig and Berlin, Germany, King’s College London, London, UK and the Department of Medical Engineering, Technical University of Berlin, Berlin, Germany 
(e-mail: tobias.schaeffter@ptb.de)}
\thanks{C. Kolbitsch is with the Physikalisch-Technische Bundesanstalt (PTB), Braunschweig and Berlin, Germany and King’s College London, London, UK
(e-mail: christoph.kolbitsch@ptb.de)}
\thanks{M. Haltmeier is with the Department of Mathematics, University of Innsbruck, Innsbruck, Austria (e-mail:markus.haltmeier@uibk.ac.at)}

}

\markboth{}
{Kofler  \MakeLowercase{\textit{et al.}}:  }

\maketitle

\begin{abstract}
In this work, we propose an iterative reconstruction scheme (ALONE - Adaptive Learning Of NEtworks) for 2D radial cine MRI based on ground truth-free unsupervised learning of shallow convolutional neural networks. The network is trained to approximate patches of the current estimate of the solution during the reconstruction. By imposing a shallow network topology and constraining the  $L_2$-norm of the learned filters, the  network's representation power is limited in order not to be able to recover noise. Therefore, the network can be interpreted to perform a low dimensional approximation of the patches for stabilizing the  inversion process. We compare the proposed reconstruction scheme to two ground truth-free reconstruction methods, namely a well known  Total Variation (TV) minimization and an unsupervised adaptive Dictionary Learning (DIC) method. The proposed method outperforms   both methods  with respect to all reported quantitative measures. Further, in contrast to  DIC, where the sparse approximation of the patches involves the solution of a complex optimization problem, ALONE only requires a forward pass of  all patches through the shallow network and therefore significantly accelerates the reconstruction.
\end{abstract}


\IEEEpeerreviewmaketitle

\section{Introduction}

Magnetic Resonance Imaging (MRI) is a widely used and indispensable medical tool for the non-invasive assessment of various diseases. For example, dynamic cardiac MRI, allows for the assessment of the cardiac function. Thereby, a certain number of cardiac phases is obtained.\\
However, MRI is well-known to suffer from relatively long acquisition times which for example limit the achievable temporal resolution which is required for a proper diagnosis. Therefore, in order to accelerate the measurement process, different techniques have emerged in field of MRI. For example, Parallel Imaging \cite{weiger2000cardiac,lin2004parallel} allows the acceleration of the data-acquisition process as solution implemented on a hardware level. In addition, to further accelerate the acquisition process, instead of acquiring the full $k$-space data in Fourier-domain, undersampling schemes have been extensively investigated in the literature. \\
Compressed Sensing theory \cite{candes2008introduction,donoho2006compressed} delivers theoretical guarantees and error bounds on the the signal-recovery from a set of random measurements. However, randomly sampling $k$-space data is challenging from a technical point of view and therefore, different undersampling schemes have been investigated in the literature, \cite{seiberlich2011improved, winkelmann2006optimal}. In particular, radial undersampling has the advantages of oversampling the center of $k$-space which contains the Fourier-coefficients corresponding to the basis functions with lower frequencies. As most physiological motions are smooth (e.g. heart contraction), radial undersampling has been reported to be particularly suitable for dynamic cine MRI applications.
When undersampling $k$-space data, the underlying inverse problem becomes an underdetermined one and no unique solution exists. Therefore, several regularization techniques have been proposed in the past, either based on hand-crafted priors, e.g.\, Total Variation (TV)-minimization \cite{block2007undersampled},  or regularizations based  on information learned from data, e.g.\ Dictionary Learning \cite{ravishankar2010mr}, \cite{caballero2014dictionary}, \cite{wang2014compressed}.\\
Recently, Neural Networks (NNs)  have been widely applied within the field of  inverse problems. Most commonly, the convolutional NNs (CNNs) are either applied as post-processing methods to reduce artefacts or denoise images, see for example \cite{jin2017deep}, \cite{han2018framing}, \cite{Hauptmann2019}, \cite{kofler2019} or employed in so-called iterative or cascaded neural networks  \cite{adler2017solving}, \cite{adler2018learned}, \cite{hammernik2018learning}, \cite{schlemper2017deep}, \cite{qin2018convolutional}. In the latter, the network architectures consist of CNNs as well as layers containing the forward and the adjoint operators which are used to ensure that the output of the CNNs match the acquired raw data. However, most methods based on CNNs nowadays are based on supervised learning (SL), i.e.\ on the implicit assumption of the availability of a large enough dataset of pairs. The literature in which CNNs using Unsupervised Learning (UL) are applied, is highly under-represented.
In this work, we propose a method for image reconstruction  in 
undersampled $2D$ radial cine MRI using an adaptive unsupervised learning approach, where the regularization is learned during the reconstruction process.
Let $\Au \colon \C^N \to \C^m$ be the undersampled dynamic radial cine MRI forward operator, $\yu \simeq  \Au \xx $ the available $k$-space data of the unknown image $\xx$. Let $\Ed_j \colon  \C^N \to \C^d$ denote the operator that extracts  the $j$-th  patch and $\net_\theta \colon \C^d \to \C^d$ a  neural network (see Section \ref{sec:problem} for precise formulations).  
The  proposed approach is based  on minimizing the functional     
\begin{multline}\label{eq:main}
\reg_{\yu,\la}(\xx, \theta) \triangleq
\frac{1}{2}\| \Au \xx - \yu \|_2^2  
\\ +  \frac{\lambda}{2} \sum_{j=1}^p \| \Ed_j (\xx) - \net_\theta \big(\Ed_j (\xx) \big) \|_2^2 
 + \pen(\theta)
\end{multline}
jointly over $\xx \in X$ and  the set of trainable parameters $\theta \in \R^q$. 
The term  $\sum_{j=1}^p \| \Ed_j (\xx) - \net_\theta\big(\Ed_j (\xx)\big) \|_2^2 $ acts as
regularizer defined by a neural network that is adapted to the specific data, i.e.\ all the available image-patches, and  $\pen$ is a penalty that  prevents overfitting of the network to noise. \\
In order to minimize \eqref{eq:main}, we  propose an iterative minimization procedure  (ALONE- Adaptive Learning Of NEtworks) that performs minimization steps in $\xx$ and $\theta$ in an alternating manner.  The update step of  the network parameters amounts to network training on patches of the current iterates. Therefore the network is trained in a completely unsupervised manner without  needing to rely on ground truth image data.    As we shall  demonstrate, ALONE  can be used with rather small patch size and shallow convolutional neural networks. As a result, ALONE  is numerically efficient, comes without any preceding training phase and  does not require artefact-free ground truth data. To the best of our knowledge, we are not aware  of any CNN-based image reconstruction  method sharing similar features.\\      
The rest of the paper is structured as follows. In Section \ref{section_Problem_Formulation}, the inverse problem as well the proposed reconstruction algorithm are formally introduced and discussed. In Section \ref{section_Experiments}, we introduce the quantitative measures which are used to evaluate the performance of our method. We compare it to the well-known TV-minimization approach, a dictionary learning-based approach and a method using previously trained CNNs to generate priors which are then used in an iterative reconstruction. We then conclude the work with a discussion and some conclusions in Section \ref{section_discussion} and \ref{section_conclusion}.
 Note that while we  focus our  presentation on dynamic radial  MRI, we point out   that the proposed framework can be used  for general 2D or 3D image reconstruction problems as well.
\section{Proposed Reconstruction  Framework}
\label{section_Problem_Formulation}
In this  section, we give  a precise  problem formulation and 
introduce the proposed unsupervised adaptive deep neural network 
based reconstruction framework. 
\subsection{Problem Formulation}
\label{sec:problem}
We consider the problem given by 
\begin{equation}\label{inv_problem}
\Au \xx = \yu,
\end{equation}
where the forward operator $\Au$ is given by the composition $\Su \circ \Ad$, of a binary mask $\Su$ and the $\Ad$ the (discretized) $2D$ frame-wise Fourier-encoding operator which samples the $k$-space data along radial lines. More precisely, the radial trajectories are chosen according to the golden-angle radial method \cite{Feng2015,feng_mrm_2012}. The coefficients are assumed to be enumerated by a set of indices $I \subset J = \{1,\ldots, N_{\mathrm{rad}}\}$ with $| I |\triangleq m < N_{\mathrm{rad}}$ which corresponds to a subset of all $N_ {\mathrm{rad}}$ Fourier coefficients that could be sampled. The number $N_{\mathrm{rad}}$ is more precisely specified by  the MR-acquisition parameters, i.e. by the number of radial trajectories, the number of receiver coils, the number of acquired cardiac phases, etc. For further details about a possible implementation of the radial Fourier-encoding operator, we refer to \cite{lin2018python}.\\
The vector $\yu \in \mathbb{C}^m$ contains the undersampled $k$-space data and the goal is to reconstruct a complex-valued 3D image $\xx \in \mathbb{C}^N$ with $N = N_x \times N_y \times N_t$ from the measurements $\yu$. Due to the application of the binary mask $\Su$, the Nyquist criterion is violated and the direct reconstruction from the measured data yields images which are contaminated by artefacts . Addressing the undersampling issue  requires the use of  proper regularization techniques which exploit  structure in the  manifolds of potential solutions to provide high quality and aliasing artefact-free results.\\
In the following, for convenience, we write 
\begin{align*}
\norm{\Ed (\xx)- \net_\theta \big( \Ed (\xx)\big)}_2^2
&\triangleq
\sum_{j=1}^p \| \Ed_j (\xx) - \net_\theta\big(\Ed_j (\xx)\big) \|_2^2 
\\
\Ed (\xx)  &\triangleq  \big(\Ed_1(\xx), \dots , \Ed_p(\xx)\big) 
\\
\net_\theta \big(\Ed (\xx)\big)  
&\triangleq  \Big(\net_\theta \big(\Ed_1(\xx)\big), \dots , \net_\theta\big( \Ed_p(\xx)\big)\Big).  
\end{align*}
Here, $\Ed_j \colon \C^N  \to \C^d$ for  $j=1, \dots, p$ is the  operator which extracts  the $j$-th 3D patch (i.e.\ a small sub-portion of the image) from an image $\xx$, $\net_\theta \colon \C^d \to \C^d $ is a neural network with trainable parameters $\theta \in \R^q$ operating on the patches. 
The number $p$ results from the shape of the patches and the strides used to extract the patches. 
As presented in the introduction, we approach  the reconstruction problem as finding a regularized solution  $\xx \in \mathbb{C}^N$ by jointly minimizing \eqref{eq:main}  over $\xx \in \mathbb{C}^N$ and  $\theta 
\in \R^q$.  Using the above introduced notation, this amounts to the 
optimization problem   
\begin{multline} \label{eq:opt}
\reg_{\yu,\la}(\xx, \theta) =
\frac{1}{2}\| \Au \xx - \yu \|_2^2  
\\ + 
\frac{\lambda}{2} \norm{\Ed (\xx)- \net_\theta\big( \Ed (\xx)\big)}_2^2  + \pen(\theta) 
\to \min_{\xx, \theta} \,.
 \end{multline}
Here,  $\pen$ denotes a  regularization imposed on the parameter set $\R^q$. It limits the capacity of the network $\net_\theta$ such that it does not adapt to image noise.  In order to minimize  \eqref{eq:main} we propose an alternating  minimization algorithm described below.

\subsection{ Proposed  Reconstruction Algorithm}

\begin{figure}
\centering
\includegraphics[width=0.95\linewidth]{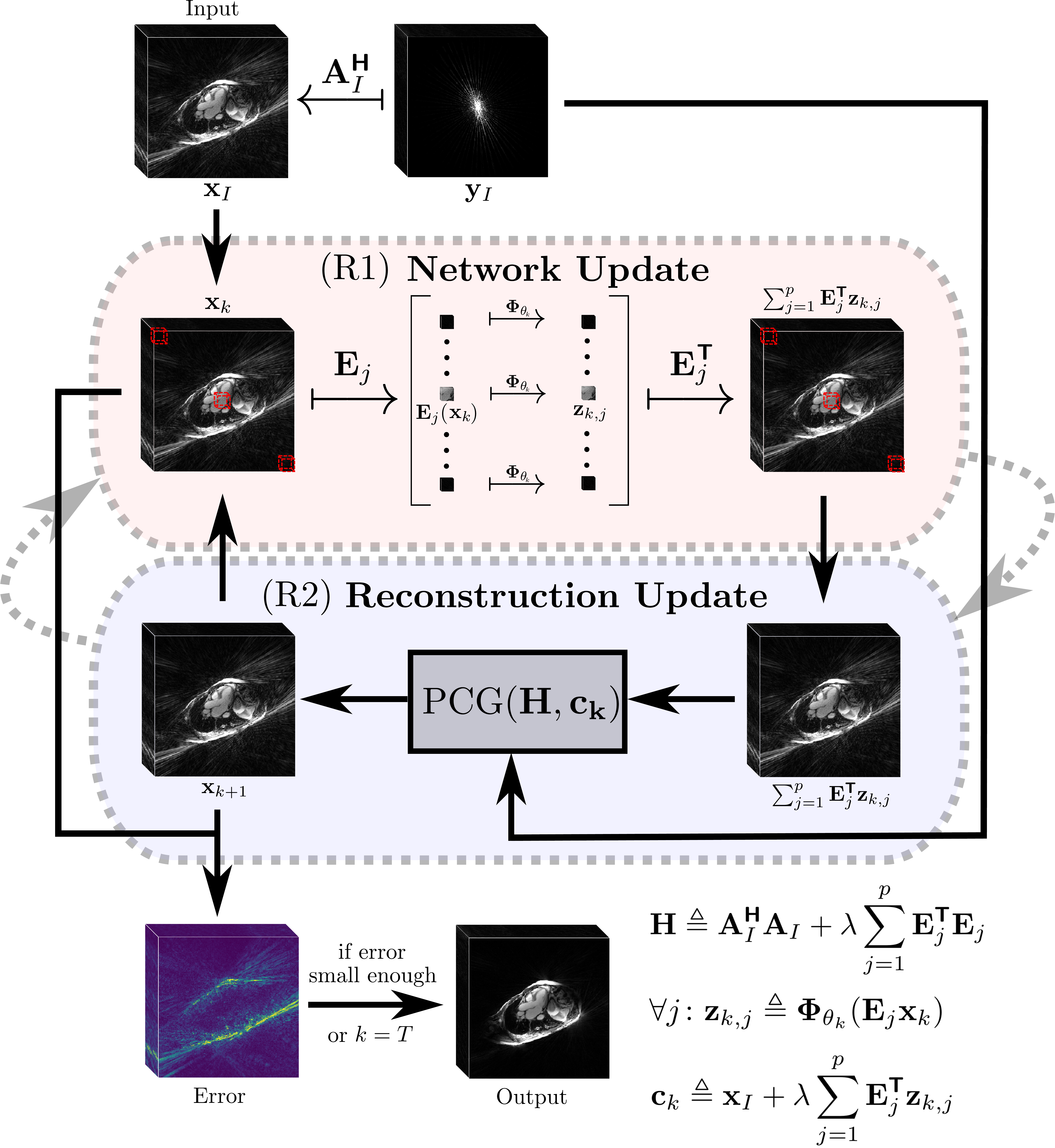}
\caption{Reconstruction algorithm ALONE. Network update: from the current image estimate $\xx_k$, 3D patches are extracted and used for training the CNN $\net_\theta$ in an unsupervised manner. Then, after training,  all patches are processed using the CNN $\net_\theta$ and reassembled to obtain a regularized solution $\net_\theta (\xx_k)$. The regularized solution is then used in a reconstruction update step, which updates the image estimate  using PCG. }\label{recon_pipeline}
\end{figure}

We begin the reconstruction process by applying the adjoint operator to the measured data and obtaining an initial guess of the solution $\xu \triangleq 
\Au^\herm \yu$. Then, we proceed  by alternating between  the following minimization steps (R1) and (R2)  with respect to $\xx$ and $\theta$, respectively.
\begin{enumerate}[label=(R\arabic*), wide]
\item \textbf{Network update:}
We first update the set of parameters $\theta \in \R^q$. For this purpose, 
we fix $\xx \in \C^N$ and solve 
\begin{multline}\label{eq_L_theta}
\fun_{\xx, \yu, \lambda}(\theta) \triangleq 
\frac{\lambda}{2} \sum_{j=1}^p \| \Ed_j (\xx) - \net_\theta\big(\Ed_j(\xx)\big) \|_2^2  \\ 
+  \pen(\theta)  \to \min_{\theta } \,.
\end{multline}
Minimizing the  loss function $\fun_{\xx, \yu, \lambda}(\theta)$ clearly  corresponds to training the network $\net_\theta$ on a dataset of pairs of patches which are extracted from the current image estimate $\xx$. The aim of the network is to reproduce the relevant (low-dimensional) information contained in the patches $\Ed_j (\xx)$ and discard the (high-dimensional) noise-like artefacts. Therefore, for each patch, the network can be interpreted to perform a low-dimensional approximation of the patches. In practice, problem (\ref{eq_L_theta}) can be efficiently solved by employing state-of-the-art non-linear optimization routines, e.g, the ADAM optimizer \cite{kingma2014adam}.
\item  \textbf{Reconstruction update:}
After having obtained an estimate for $\theta$, we set $\zz_j 
= \net_\theta \big(\Ed_j (\xx) \big)  \in \C^d$ for  any patch and and update the image estimate 
$\xx \in \C^N$  by solving  
\begin{multline}\label{eq_L_x}
\fun_{\theta, \yu, \lambda} (\xx) \triangleq \frac{1}{2}\| \Au \xx - \yu \|_2^2 
\\
+ \frac{\lambda}{2} \sum_{j=1}^p 
\norm{\Ed_j (\xx) - \zz_j}_2^2 \rightarrow \min_{\xx}   \,.
\end{multline}
The optimization  problem  \eqref{eq_L_x} is quadratic  and hence   
can be solved efficiently. More precisely, according to Fermat's rule, 
$\xx$ solves  \eqref{eq_L_x} if and only if  it satisfies the linear optimality 
 condition  
\begin{equation}
\mathbf{0} = 
\nabla_\xx \fun_{\theta, \yu,\lambda} (\xx) 
=
 \Hd \xx - \cc \,,
\end{equation}
with
 \begin{align}
\Hd &\triangleq \Au^\herm \Au + \lambda \sum_{j=1}^p \Ed_j^\trans  \Ed_j, \label{Hop} \\ 
\cc  &\triangleq  \xu + \lambda \sum_{j=1}^p \Ed_j^\trans (\zz_j). \label{ck}
\end{align}
If the  operator $\Ad$ is an isometry, e.g. when the full data acquisition takes place using a single-coil and sampling along a Cartesian grid, it holds $\| \Ad \xx\|_2 = \|\xx\|_2$ for all $\xx$ and, consequently, problem (\ref{eq_L_x}) has an analytic solution. It is given by performing a linear combination of the available $k$-space data $\yu$ and the one estimated by CNN-approximation and then subsequently applying the inverse operator, i.e.\
\begin{equation}\label{sol_iso}
\xx^{\ast} = \Ad^\herm \Big( \frac{\lambda}{1+ \lambda} \yu + \mathbf{\Lambda} \Ad\, \sum_{j=1}^p \Ed_j^\trans (\zz_j) \Big), 
\end{equation}
where the diagonal operator $\mathbf{\Lambda}$ accounts for proper weighting of the $k$-space data; see \cite{ravishankar2010mr} for a detailed derivation of \eqref{sol_iso}. In the general case, the solution of problem (\ref{eq_L_x}) can be obtained by solving the linear matrix equation  $\Hd \xx = \cc$.
The system $\Hd \xx = \cc$ can be  efficiently solved by  means  of 
any iterative scheme. Due to the symmetric structure of the operator $\Hd$, we can apply, for example,  the pre-conditioned conjugate gradient method (PCG) \cite{hestenes1952methods}.
\end{enumerate}
After having obtained the solutions to problem (\ref{eq_L_theta}) and \eqref{eq_L_x}, we repeat the procedure until a pre-defined stopping criterion is fulfilled. Let $(\xx_k)_{k\in \mathbb{N}}$ be the sequence of reconstructions obtained as just described. We stop the iteration  either if the relative change of the newly obtained solution $\xx_{k+1}$ is small enough, i.e. $\| \xx_{k+1} - \xx_k\|_2^2 / \| \xx_k\|_2^2 < \varepsilon$ for some $\varepsilon \geq 0$ or if a chosen maximal number of iterations $T>0$ has been performed. Algorithm \ref{reco_algo} summarizes the just described steps, which we name  ALONE (Adaptive  Learning Of NEtworks) reconstruction algorithm. Figure \ref{recon_pipeline} shows an illustration of the Algorithm.
\begin{algorithm}[!t]
    \caption{{Proposed ALONE algorithm}}\label{reco_algo}
  \begin{algorithmic}[1]
  \INPUT Initialization $\xx_0 = \Au^\herm \yu$
  \PARAMETER $\lambda>0$, iteration number $T>0$, accuracy $\varepsilon \geq 0$
   \OUTPUT reconstructed image $\xx_{\mathrm{reco}}$ 
  \STATE $k\gets 0$
  \STATE $e_k \gets \infty$
    \WHILE{$k \leq T$ and $e_k > \varepsilon$ }
      \STATE $\theta_{k} \gets \argmin_{\theta} \fun_{\xx_k,\yu,\lambda}(\theta) $ 
      \STATE $\forall j \colon  \zz_{k,j} \gets \net_{\theta_{k}}\big(\Ed_j (\xx_k)\big)$
      \STATE $\cc_k \gets \xu + \lambda \sum_{j=1}^p \Ed_j^\trans (\zz_{k,j})$
            
      \STATE $\xx_{k+1} \gets \argmin_{\xx}  \fun_{\theta_k,\yu,\lambda}(\xx) $
       by solving  $\Hd \xx = \cc_k$
      \STATE $e_{k+1} \gets \| \xx_{k+1} - \xx_k\|_2^2 / \| \xx_k\|_2^2$
      \STATE $k \gets k+1$
    \ENDWHILE
   \STATE  $\xx_{\mathrm{reco}} \gets  x_k$ 
  \end{algorithmic}
\end{algorithm}

\section{Experiments}\label{section_Experiments}

\subsection{Dataset and Evaluation Metrics}

For the evaluation of our proposed method, we used a dataset of four patients for which we acquired $2D$ cine MRI image sequences. The image sequences have an in-plane number of pixels of $N_x \times N_y = 320 \times 320$ and a number of cardiac phases of $N_t=30$. For two of the patients, $N_z=12$ different slices were obtained, while for the resting two patients, we could only acquire $N_z=6$ slices due to restricted respiratory capabilities. Thus, our dataset consists of a set of 36 two-dimensional cine MR image sequences. In order to quantitatively assess the performance of our method, we reconstructed all the $2D$ cine MR images using $kt$-SENSE \cite{Tsao2003, feng_mrm_2012} using $N_{ \varphi}=3400$ radial lines. Then, from the obtained image sequences we retrospectively generated radially acquired $k$-space data by only sampling along $N_{\varphi}=1130$ radial trajectories. More precisely, in our case, the operator $\Au$ is given by $\Au = \mathrm{diag}(\Su,\ldots,\Su) \circ \mathrm{diag}(\Ad,\ldots,\Ad) \circ [\mathbf{C}_1,\ldots,\mathbf{C}_{n_c}]^\trans$, where  $\mathbf{C}_i$ is the $i$-th coil-sensitivity map and $n_c=12$ and, again, $\Ad$ is the frame-wise radial Fourier encoding operator. Note that due to the used radial sampling pattern, sampling along $N_{\varphi}=3400$ spokes already corresponds to an undersampling factor of $\sim 3$. Therefore, acquiring $k$-space data along only $N_{ \varphi}=1130$ corresponds to an acceleration factor of $\sim 9$. \\
We assessed the quality of our obtained reconstructions by comparing them to the $kt$-SENSE reconstructions obtained using $N_{\varphi}=3400$ radial spokes. For the evaluation, we used the following quantitative measures: peak signal-to-noise ratio (PSNR), normalized root mean squared error (NRMSE), the structural similarity index measure (SSIM) and the Haar wavelet-based perceptual similarity measure (HPSI) \cite{reisenhofer2018haar}.
Since the field of view is quite large and image sequences contain a noticeable portion of background which is irrelevant for diagnostic purposes, before calculating the statistics, we cropped all the image sequences to $N_x \times N_y \times N_t = 160 \times 160 \times 30$ using a symmetric cut-off of 80 in $x$- and $y$-direction. 

\subsection{Network Architecture and Training}
\begin{figure}[!h]
\centering
\begin{overpic}[width=1\linewidth,tics=10]{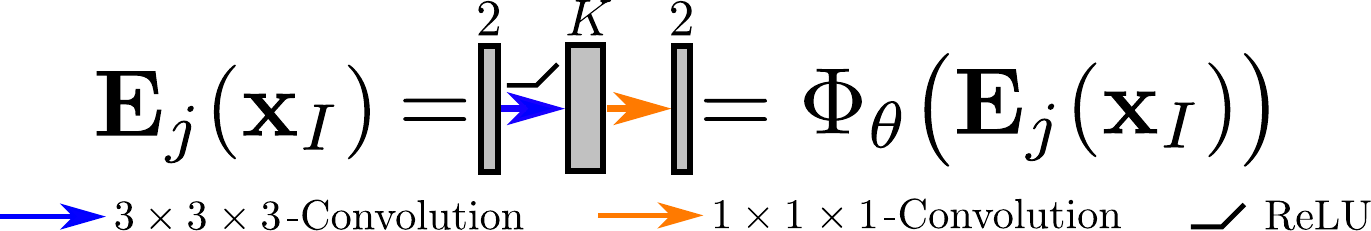}
\end{overpic}
\caption{The shallow network used in the experiments. The input is a complex-valued $3D$ patch which is extracted from the image sequence. Since the data is complex-valued, we use two channels to represent real- and imaginary part, respectively. The number of learned filters is $K$.}
\label{shallow_CNN}
\end{figure}

Here, we briefly describe the network architecture used for all the experiments. The CNN is shown in Figure \ref{shallow_CNN}. It consists of a three-layers CNN with only one hidden layer. The input of the CNN is a patch $\Ed_j(\xx)$ which is extracted from the current estimate of the image. Since the images are complex-valued we represent the patches using two-channels. The image patch is passed through a $3\times 3\times 3$-convolutional layer with $K$ filters, followed by a voxel-wise application of the ReLU activation function. Then, from the feature maps a complex-valued patch is obtained by applying a $1\times 1\times 1$-convolutional layer with the identity as activation function. Therefore, the output patch corresponds to a learned linear combination of the extracted $K$ feature maps which are learned by the $K$ filters.\\
Intuitively speaking, the network $\net_\theta$ is trained to perform a learned  dimensionality reduction of the $3D$ patches which are supposed to lie on a lower dimensional manifold. Similar to dictionary learning, where signals are represented as sparse combinations of elements of an overcomplete basis, our method performs a dimensionality reduction representing each $3D$ patch as a linear combination of last extracted feature maps which depend on the learned $K$ filters. However, in contrast to dictionary learning, where, once the dictionary is learned, the correspondent support of the signals has to be calculated by some sparse coding algorithm, our method extracts $K$ filters which can be globally used for all the patches.\\
Since the network $\net_\theta$ is trained in an unsupervised manner on the patches of the current image estimate, the network's representation power has to be constrained in order make it a proper regularization. The first restriction is directly given by the fact that the network is very shallow and only contains one hidden layer. Second, the number of learned filters is chosen to be  quite small, for example $K=16$. Further, while training the network, a further regularization $\pen(\theta)$ is included in the loss function. We choose to bound the $L_2$-norm of the learned kernels, i.e. $\pen(\theta) = \sum_{k=1}^K \| f_k\|_2^2$, where $f_k$ is the $k$-th convolutional filter.

For the following experiments, we used a total number of $T=25$ iterations of ALONE. For each iteration in  the ALONE  algorithm \ref{reco_algo}, the number of back-propagations for training the network $\net_\theta$ to learn to patch-wise approximate the current image estimate was $400$. The patch-size was chosen to be $32\times 32 \times 4$. Network training was carried out by minimizing the loss function $\fun_{\xx, \yu,\lambda}(\theta)$ using the Adam optimizer \cite{kingma2014adam} with a learning rate of 0.001. Further, before training, all input patches were normalized by subtracting the mean and dividing the patch by the standard deviation. For the reassembling of the patches, the normalization is reversed after having processed them with the network $\net_\theta$. We set the number of learned filters to $K=16$ and used the $L_2$-norm of the learned kernels as parameter regularization $\pen(\theta)$ used in (\ref{eq_L_theta}). When given $\cc_k$, the system $\Hd \xx = \cc_k$ was solved by performing $n_{\mathrm{iter}}=4$ iterations of PCG.

\subsection{Comparison to Other Iterative Methods}
\begin{figure*}[!h]				
\begin{minipage}{\linewidth}
\centering
\resizebox{\linewidth}{!}{
\includegraphics[height=1.9cm]{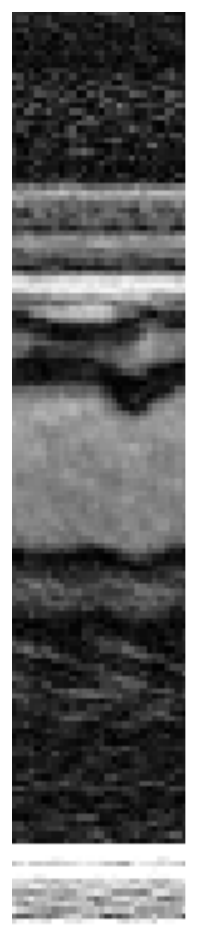}\hspace{-0.2cm}
\begin{overpic}[height=1.9cm,tics=10]{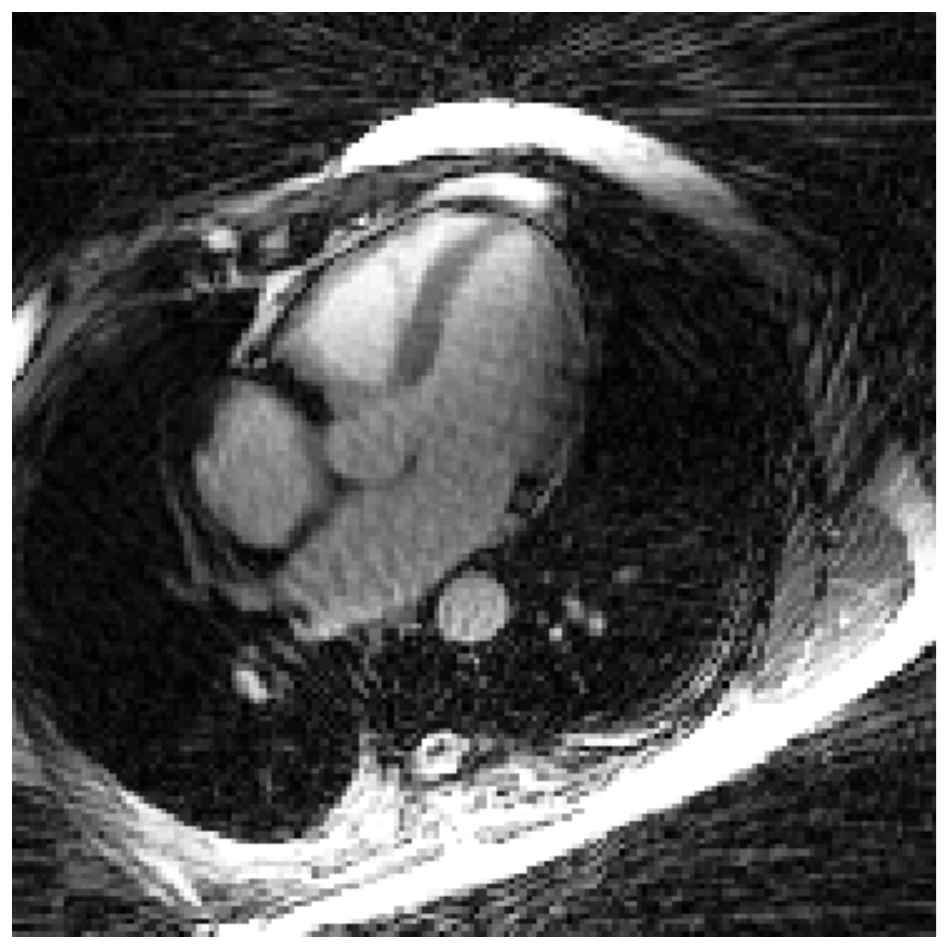}
 \put (76,78) {\small\textcolor{white}{(a)}}
\end{overpic}\hspace{-0.1cm}
\includegraphics[height=1.9cm]{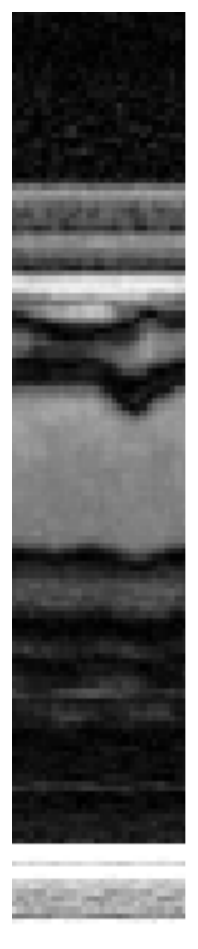}\hspace{-0.2cm}
\begin{overpic}[height=1.9cm,tics=10]{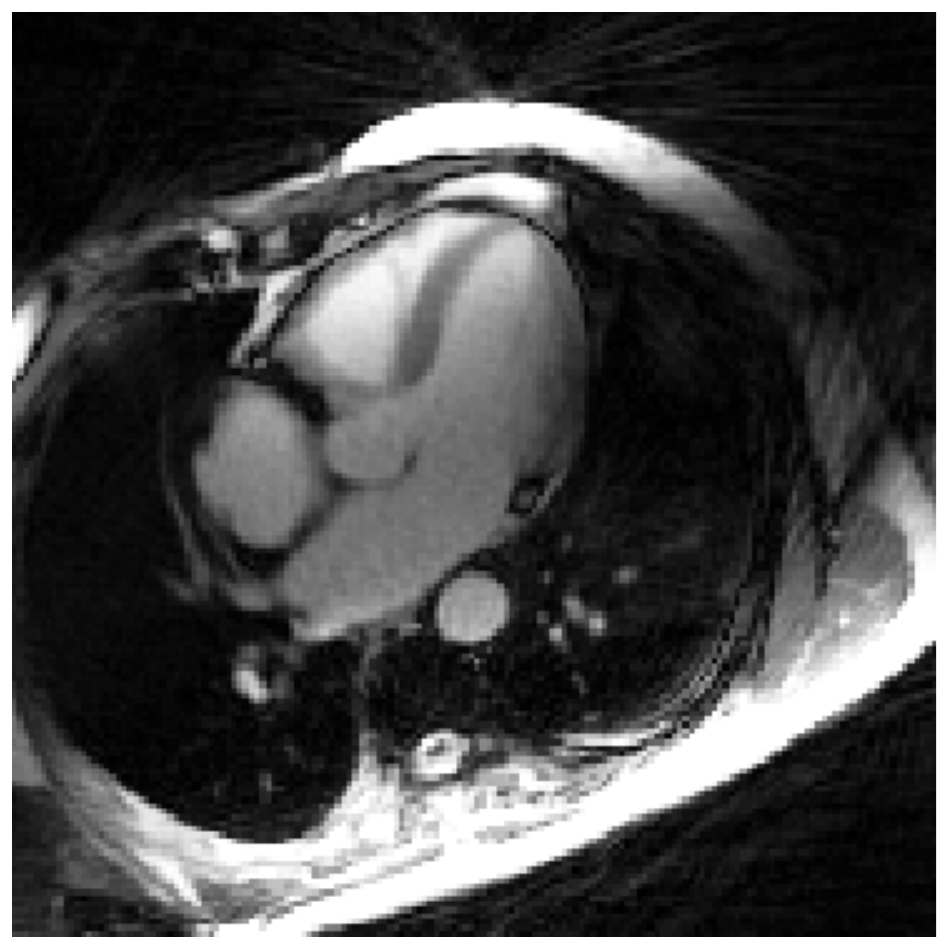}
 \put (76,78) {\small\textcolor{white}{(b)}}
\end{overpic}\hspace{-0.1cm}
\includegraphics[height=1.9cm]{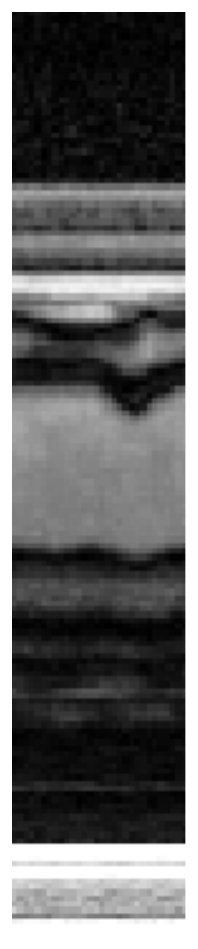}\hspace{-0.2cm}
\begin{overpic}[height=1.9cm,tics=10]{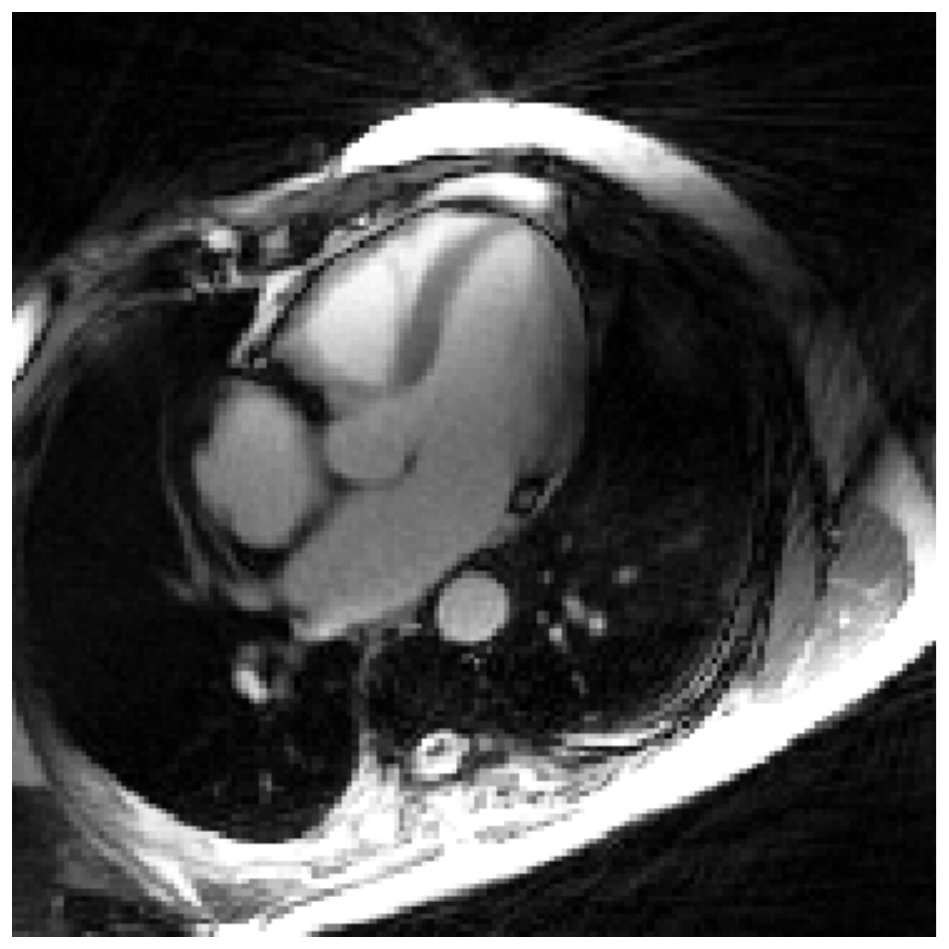}
 \put (76,78) {\small\textcolor{white}{(c)}}
\end{overpic}\hspace{-0.1cm}
\includegraphics[height=1.9cm]{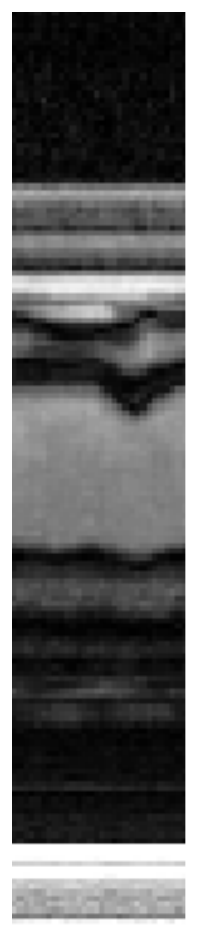}\hspace{-0.2cm}
\begin{overpic}[height=1.9cm,tics=10]{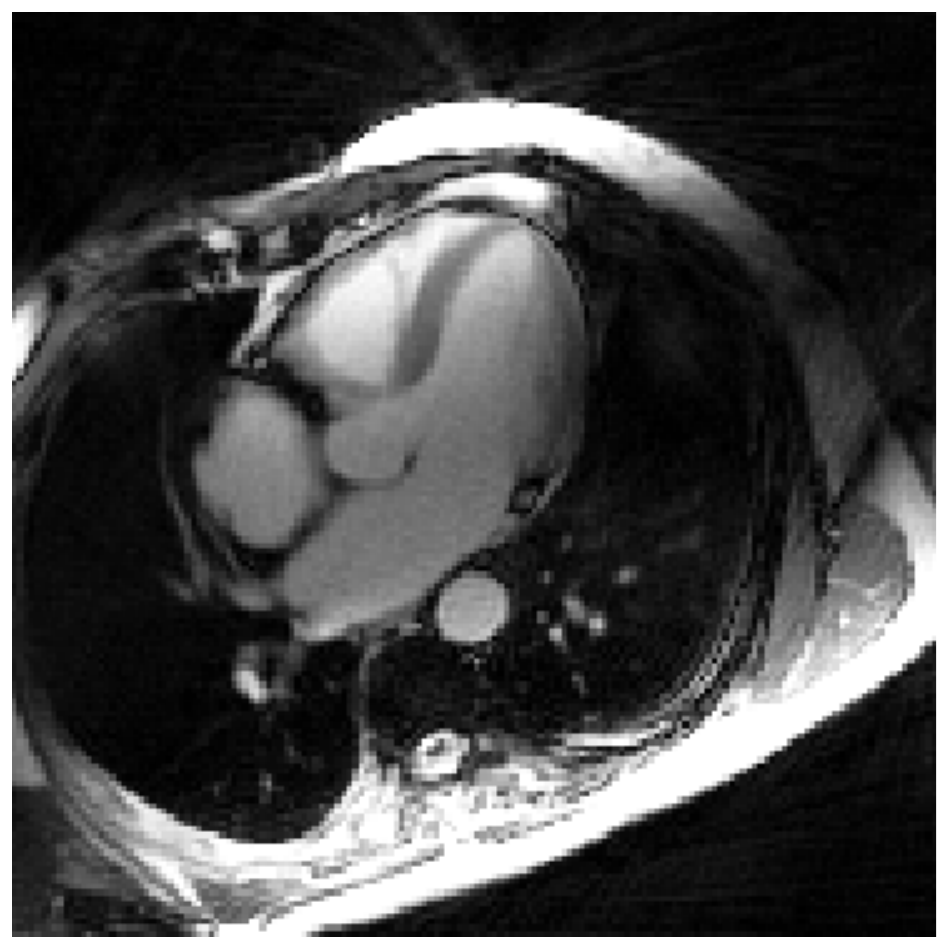}
 \put (76,78) {\small\textcolor{white}{(d)}}
\end{overpic}\hspace{-0.1cm}
\includegraphics[height=1.9cm]{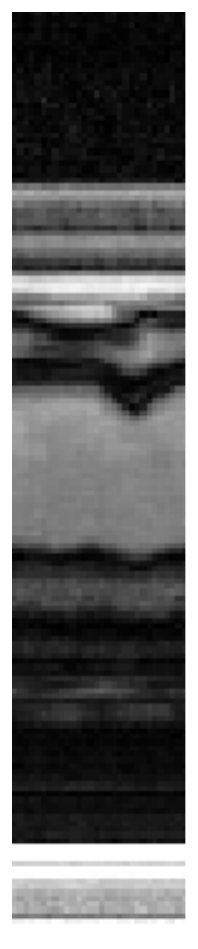}\hspace{-0.2cm}
\begin{overpic}[height=1.9cm,tics=10]{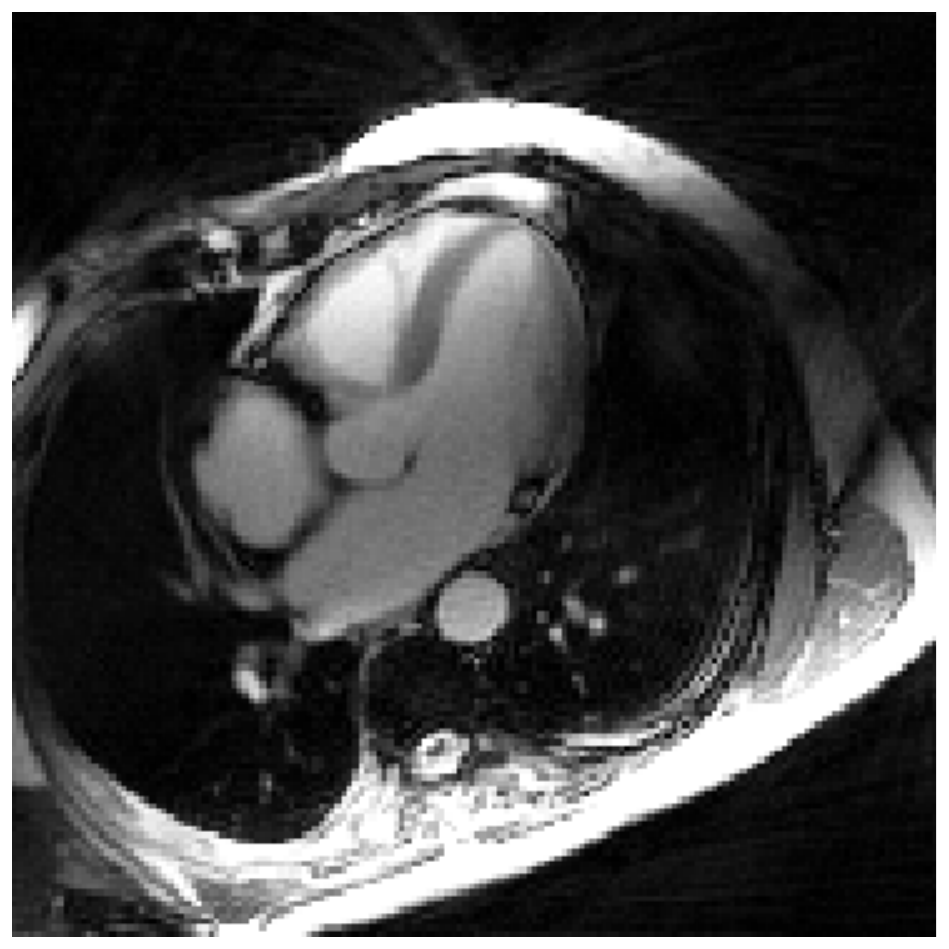}
 \put (76,78) {\small\textcolor{white}{(e)}}
\end{overpic}\hspace{-0.1cm}
}\\
\resizebox{\linewidth}{!}{
\includegraphics[height=1.9cm]{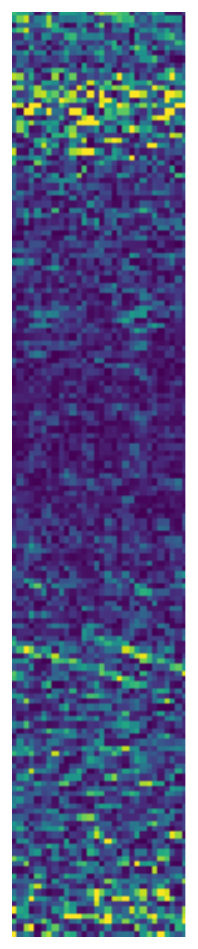}\hspace{-0.2cm}
\begin{overpic}[height=1.9cm,tics=10]{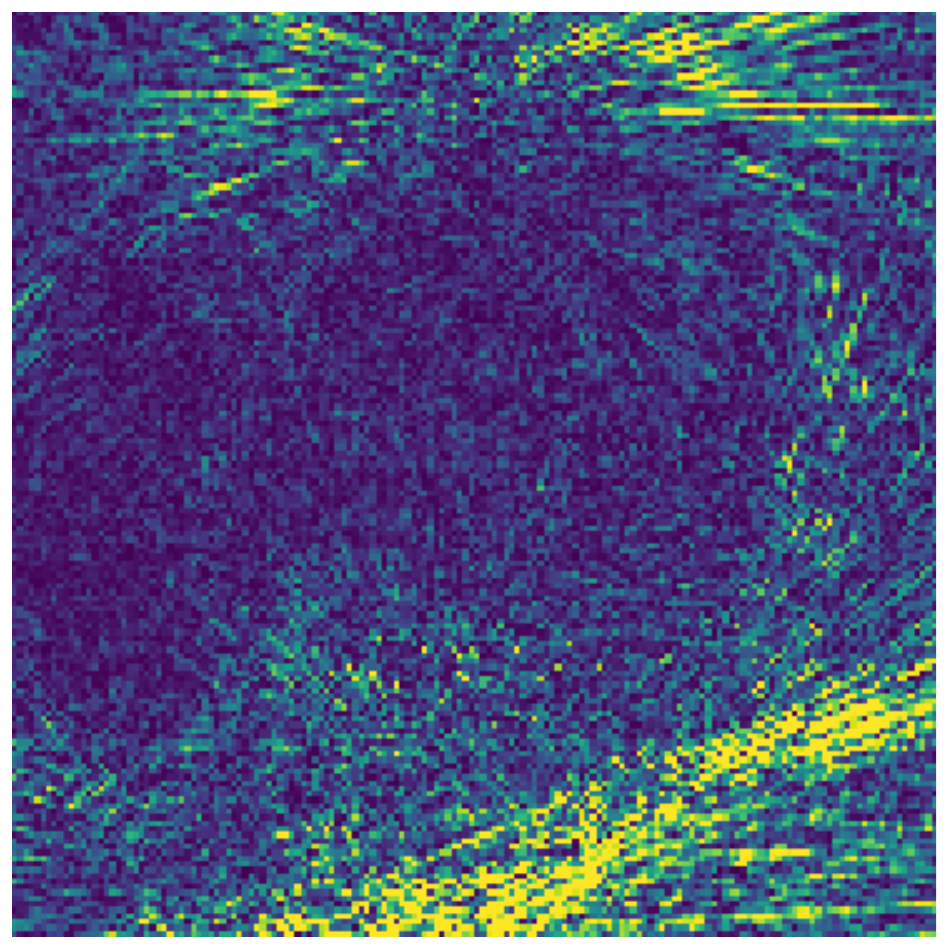}
\end{overpic}\hspace{-0.1cm}
\includegraphics[height=1.9cm]{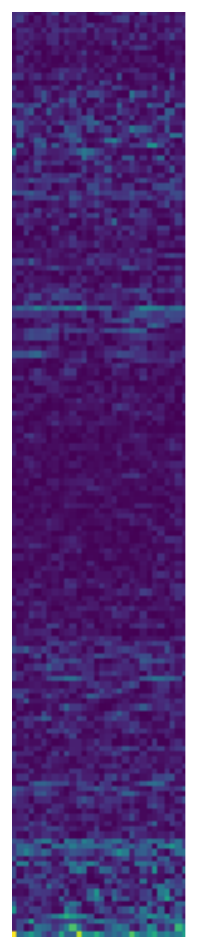}\hspace{-0.2cm}
\begin{overpic}[height=1.9cm,tics=10]{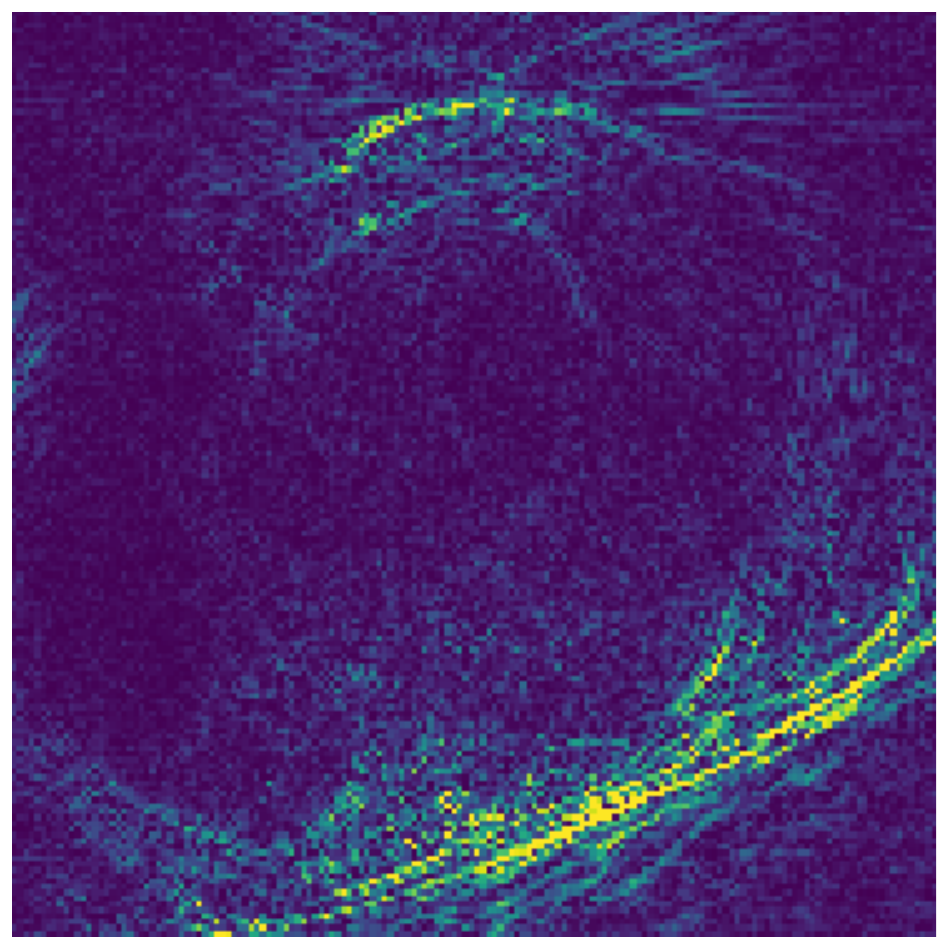}
\end{overpic}\hspace{-0.1cm}
\includegraphics[height=1.9cm]{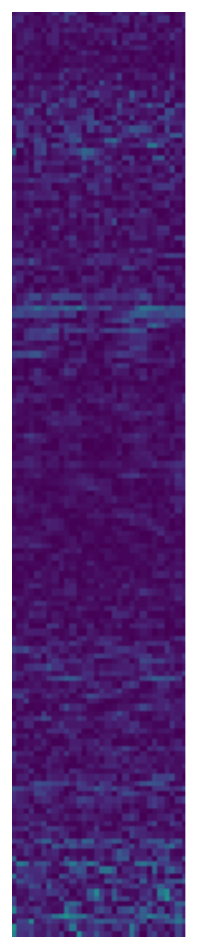}\hspace{-0.2cm}
\begin{overpic}[height=1.9cm,tics=10]{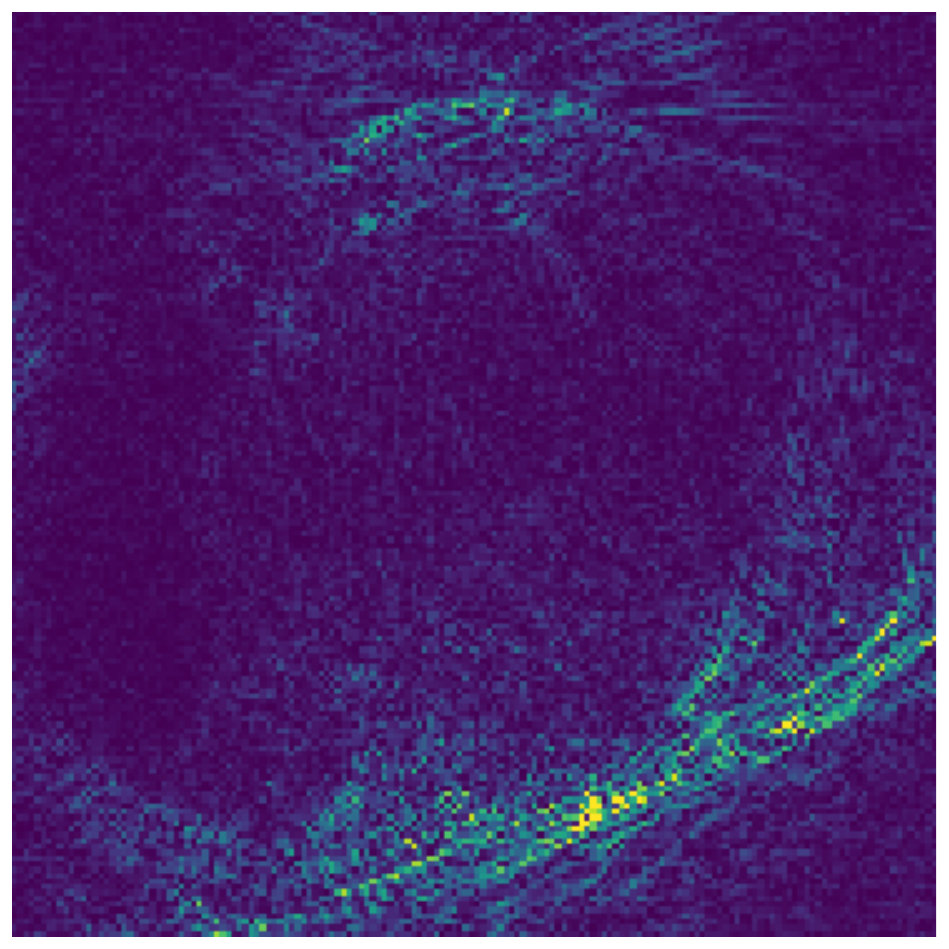}
\end{overpic}\hspace{-0.1cm}
\includegraphics[height=1.9cm]{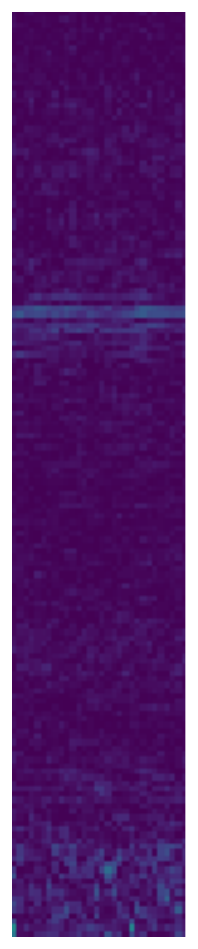}\hspace{-0.2cm}
\begin{overpic}[height=1.9cm,tics=10]{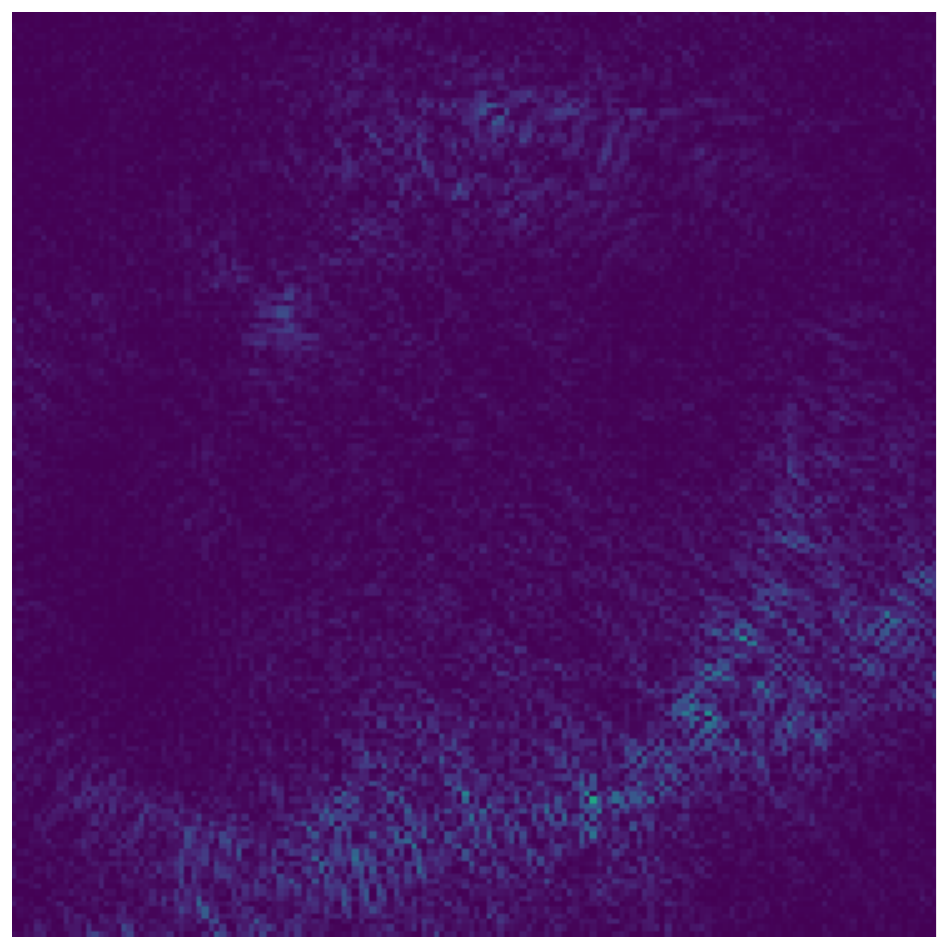}
\end{overpic}\hspace{-0.1cm}
\includegraphics[height=1.9cm]{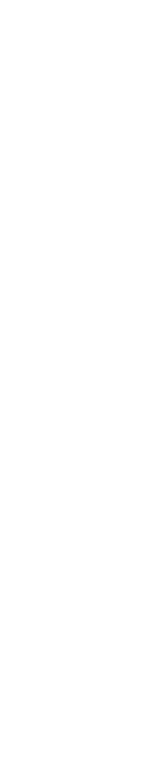}\hspace{-0.2cm}
\begin{overpic}[height=1.9cm,tics=10]{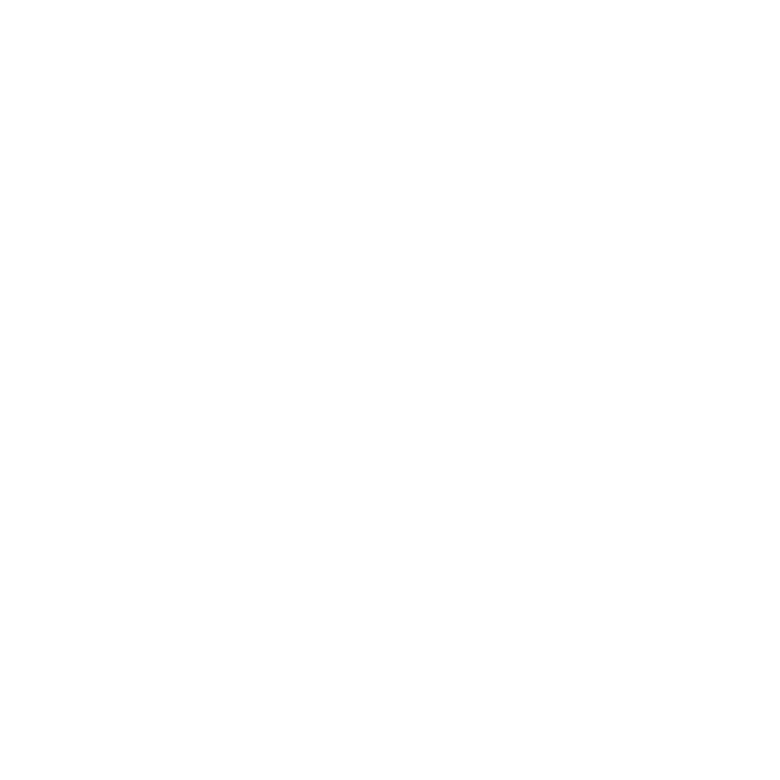}
\end{overpic}\hspace{-0.1cm}
}\\
\resizebox{\linewidth}{!}{
\includegraphics[height=1.9cm]{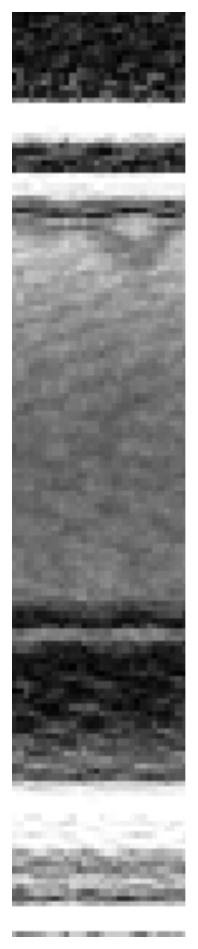}\hspace{-0.2cm}
\begin{overpic}[height=1.9cm,tics=10]{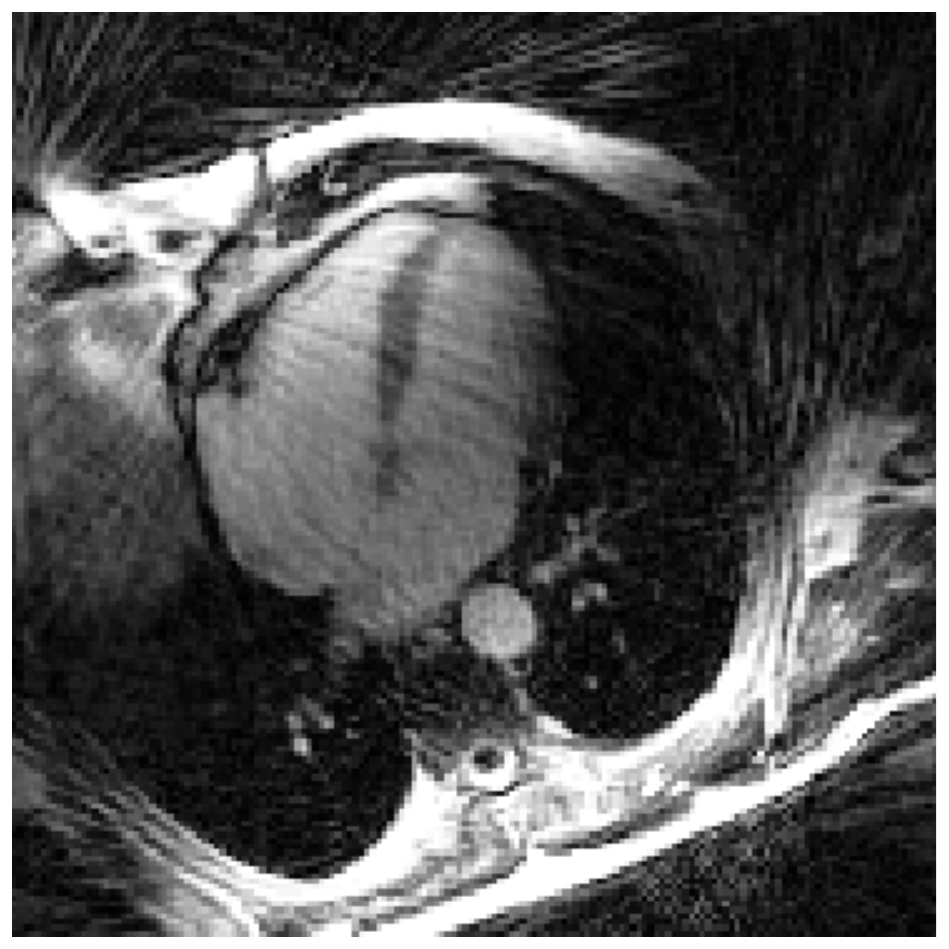}
 \put (76,78) {\small\textcolor{white}{(a)}}
\end{overpic}\hspace{-0.1cm}
\includegraphics[height=1.9cm]{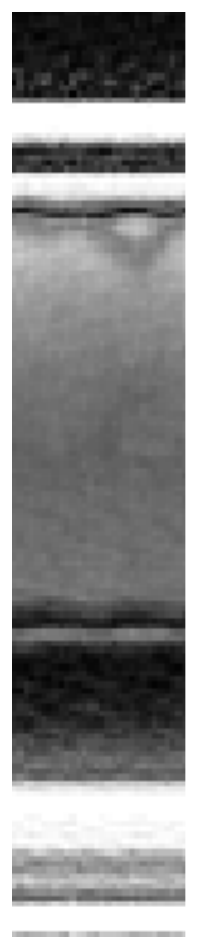}\hspace{-0.2cm}
\begin{overpic}[height=1.9cm,tics=10]{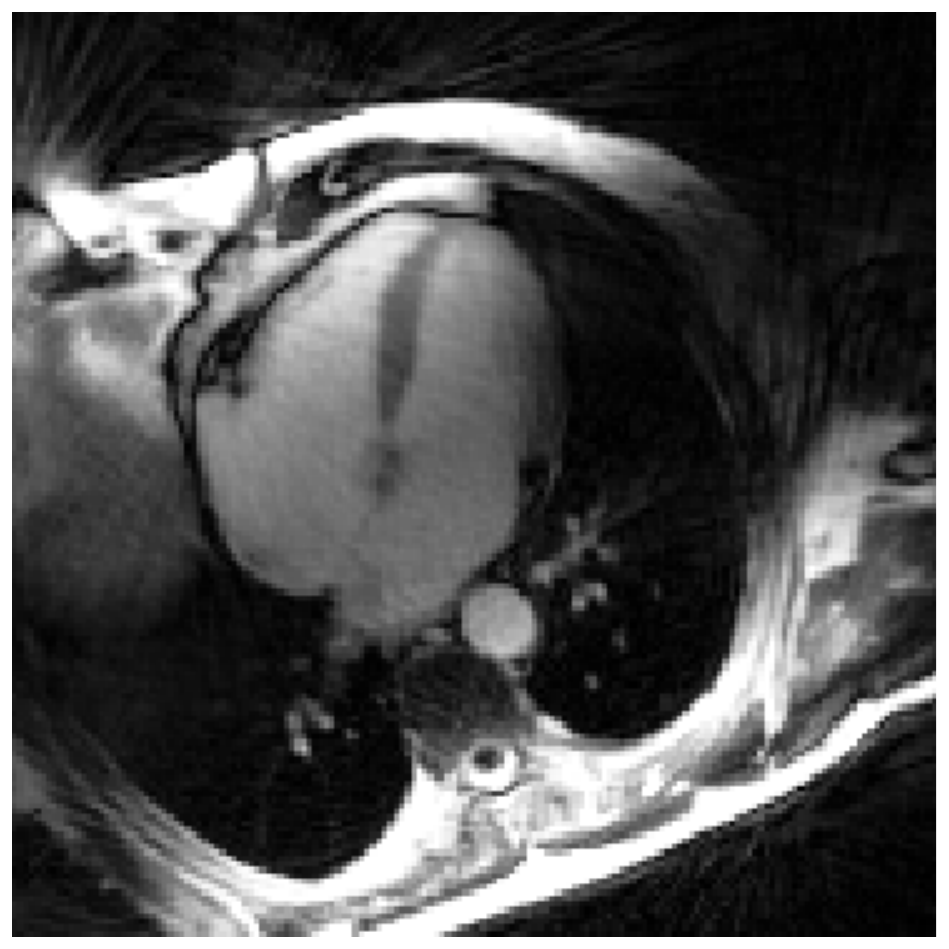}
 \put (76,78) {\small\textcolor{white}{(b)}}
\end{overpic}\hspace{-0.1cm}
\includegraphics[height=1.9cm]{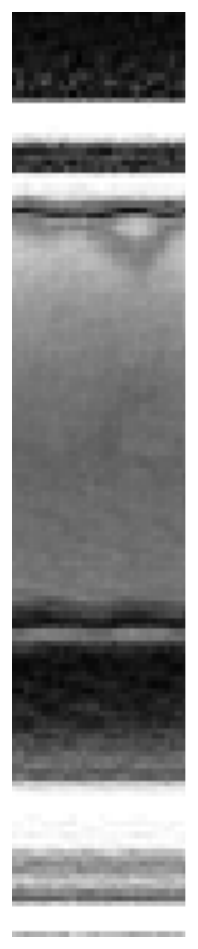}\hspace{-0.2cm}
\begin{overpic}[height=1.9cm,tics=10]{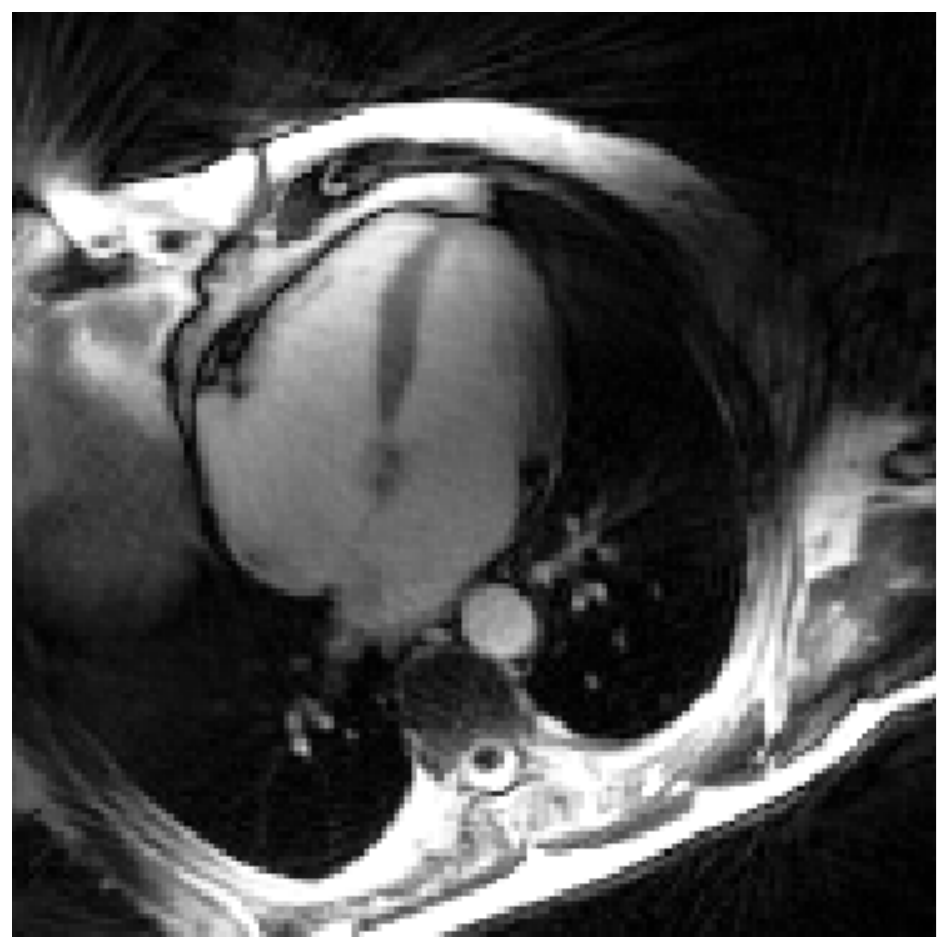}
 \put (76,78) {\small\textcolor{white}{(c)}}
\end{overpic}\hspace{-0.1cm}
\includegraphics[height=1.9cm]{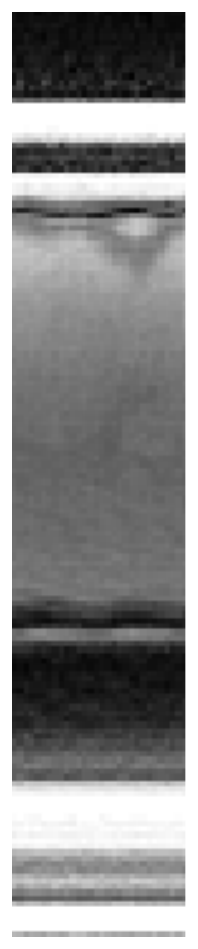}\hspace{-0.2cm}
\begin{overpic}[height=1.9cm,tics=10]{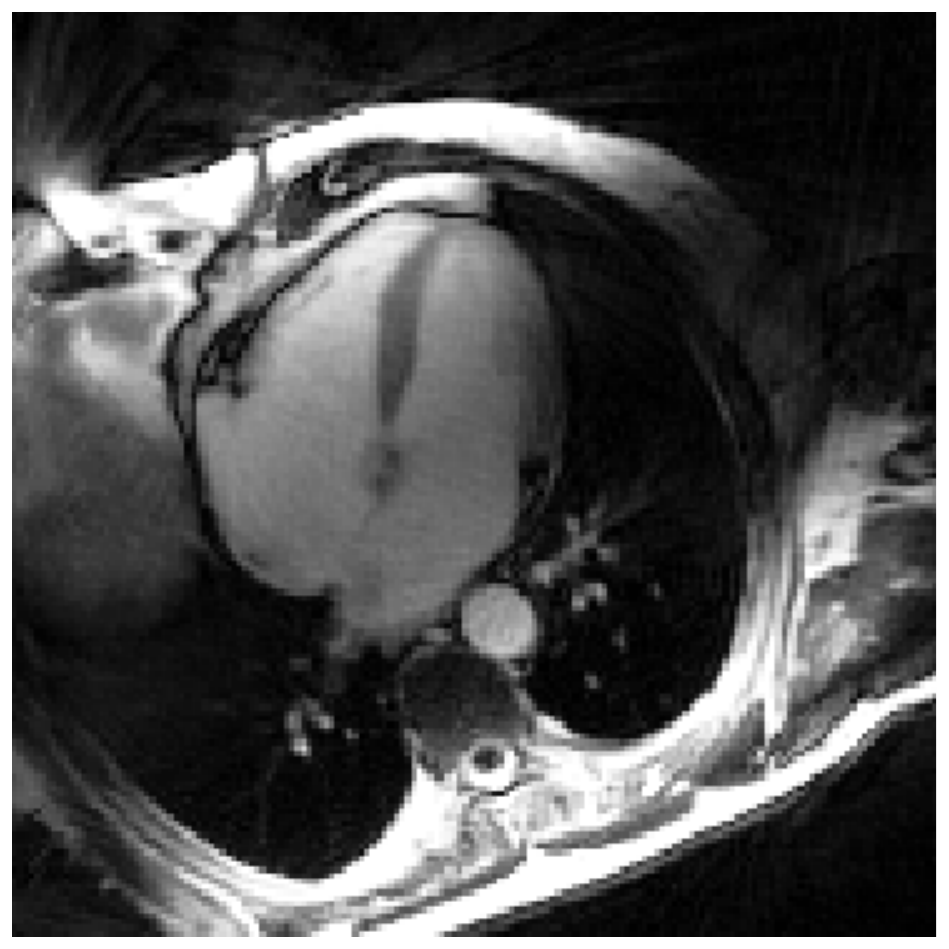}
 \put (76,78) {\small\textcolor{white}{(d)}}
\end{overpic}\hspace{-0.1cm}
\includegraphics[height=1.9cm]{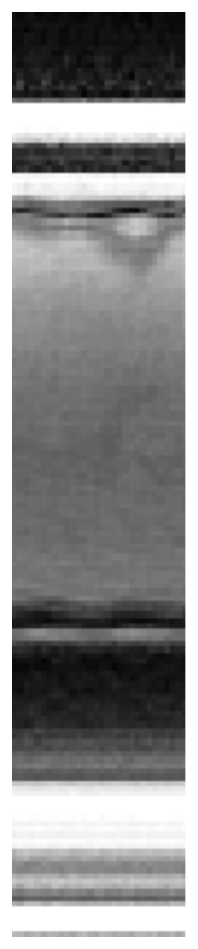}\hspace{-0.2cm}
\begin{overpic}[height=1.9cm,tics=10]{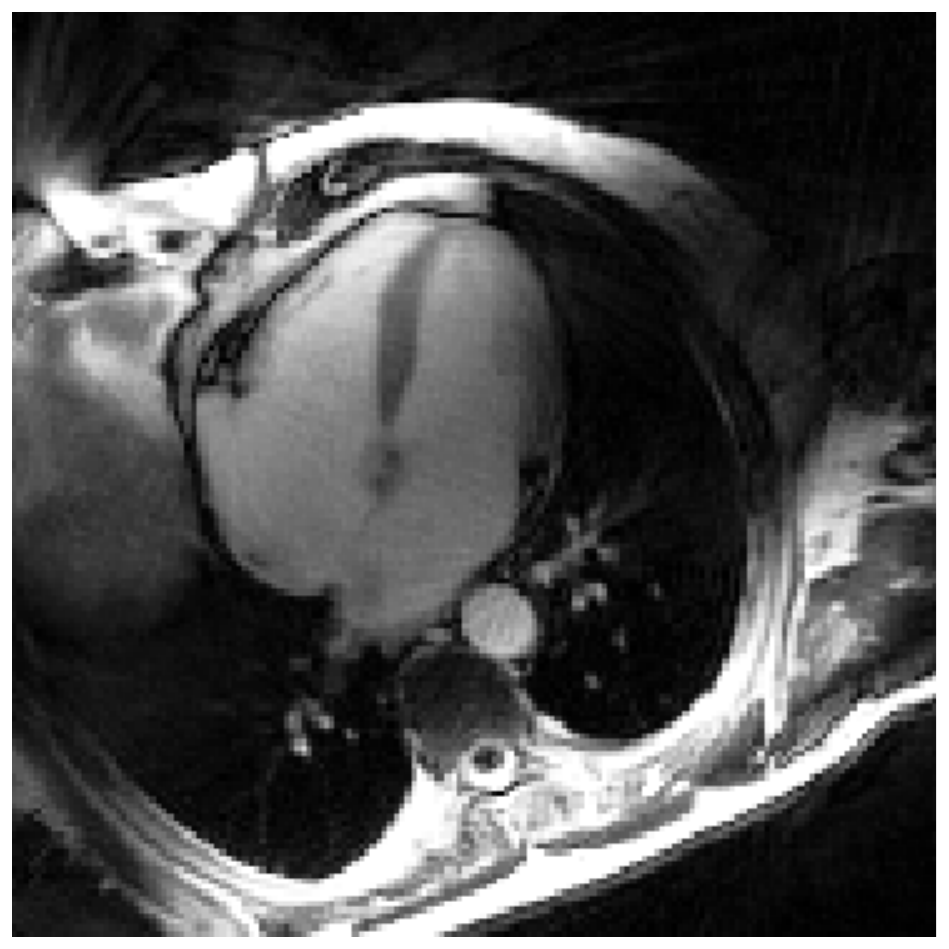}
 \put (76,78) {\small\textcolor{white}{(e)}}
\end{overpic}\hspace{-0.1cm}
}\\
\resizebox{\linewidth}{!}{
\includegraphics[height=1.9cm]{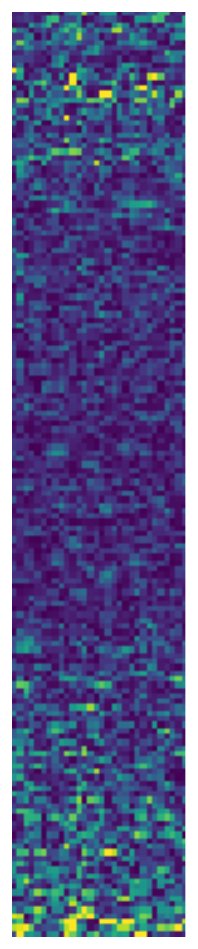}\hspace{-0.2cm}
\begin{overpic}[height=1.9cm,tics=10]{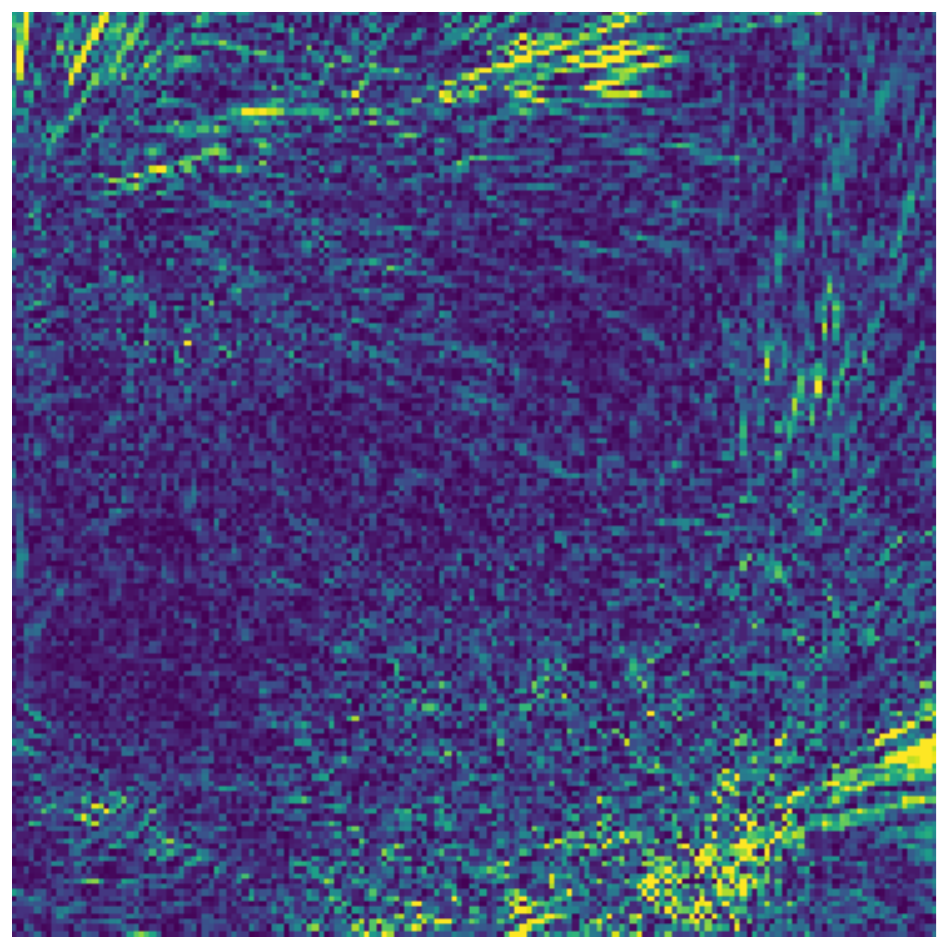}
\end{overpic}\hspace{-0.1cm}
\includegraphics[height=1.9cm]{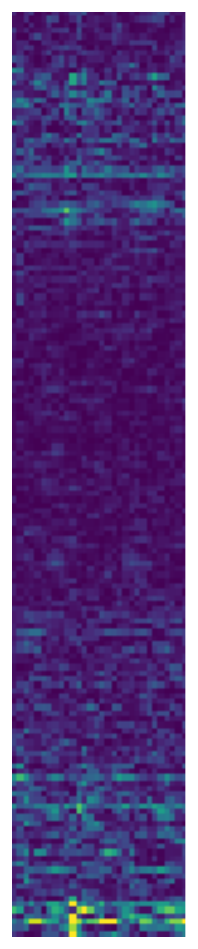}\hspace{-0.2cm}
\begin{overpic}[height=1.9cm,tics=10]{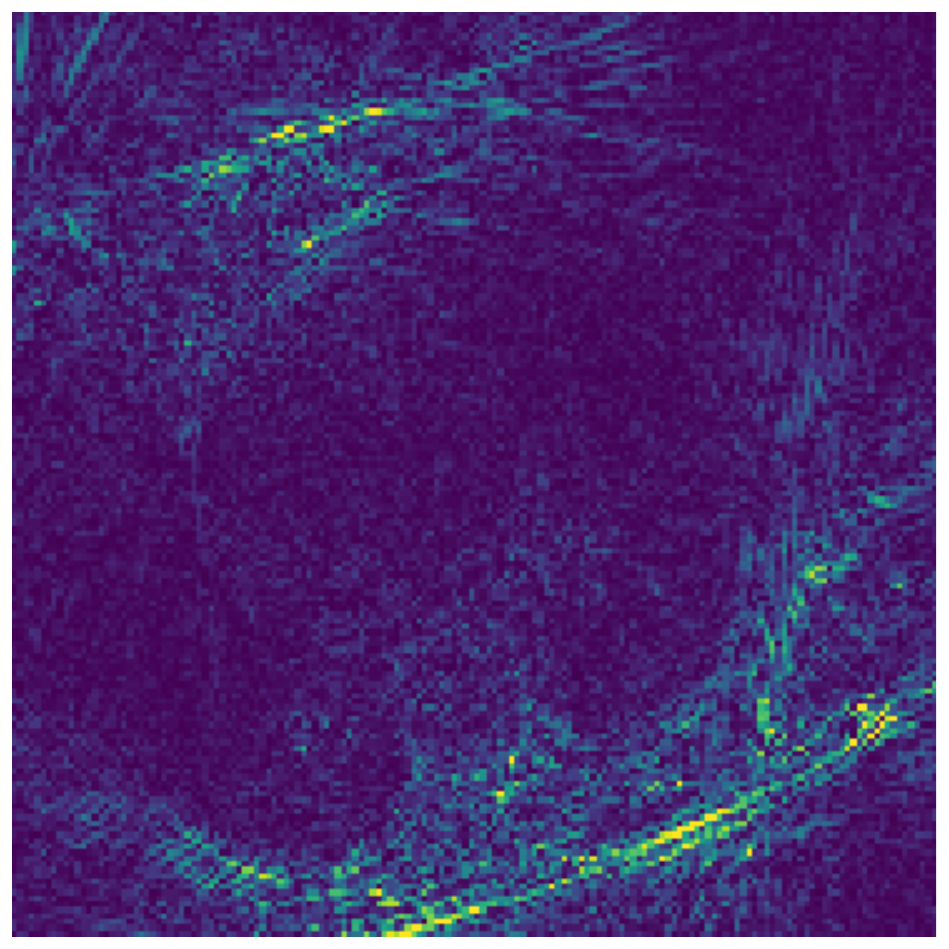}
\end{overpic}\hspace{-0.1cm}
\includegraphics[height=1.9cm]{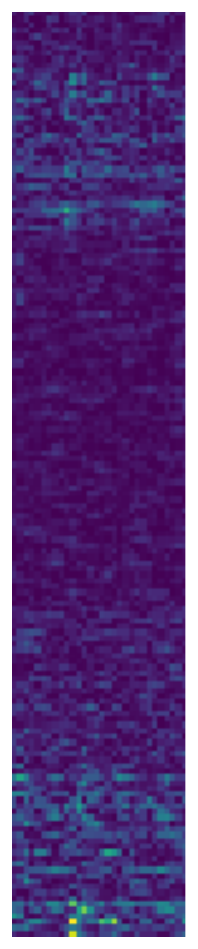}\hspace{-0.2cm}
\begin{overpic}[height=1.9cm,tics=10]{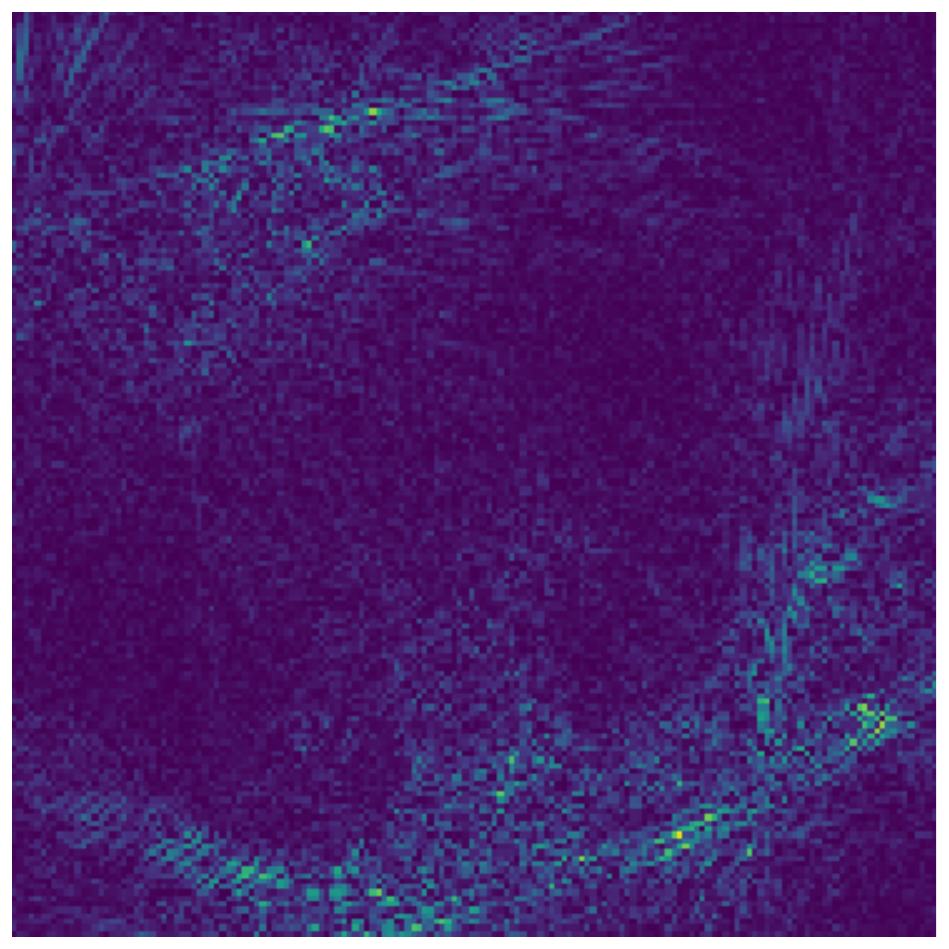}
\end{overpic}\hspace{-0.1cm}
\includegraphics[height=1.9cm]{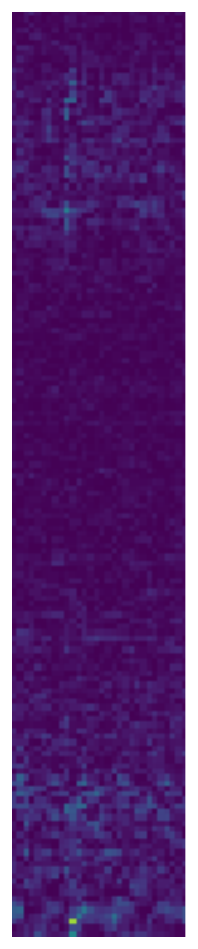}\hspace{-0.2cm}
\begin{overpic}[height=1.9cm,tics=10]{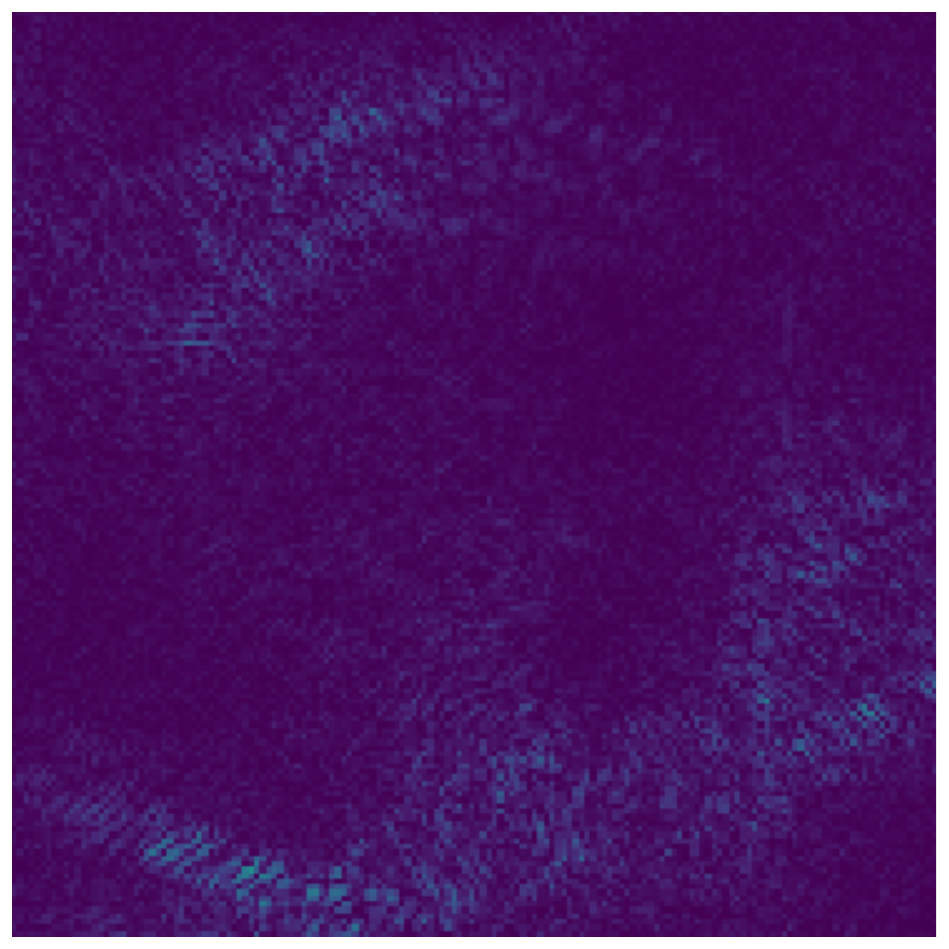}
\end{overpic}\hspace{-0.1cm}
\includegraphics[height=1.9cm]{images/results/white_images/white_xt_yt.pdf}\hspace{-0.2cm}
\begin{overpic}[height=1.9cm,tics=10]{images/results/white_images/white_xy.pdf}
\end{overpic}\hspace{-0.1cm}
}\\
\end{minipage}
\caption{Results obtained by different methods based on iterative reconstruction. NUFFT reconstruction from $N_{ \varphi}=1130$ radial spokes (a), TV minimization-based regularization (b), DIC-based regularization (c), proposed regularization based on shallow CNNs (d), $kt$-SENSE reconstruction from $N_{ \varphi}=3400$ radial spokes (e).} \label{quantitative_comparisons}
\end{figure*}

In this Section, we compare our proposed reconstruction method to the well known total variation-minimization algorithm \cite{chambolle2005total} which has been successfully applied to $2D$ cine MRI \cite{block2007undersampled} and to an iterative reconstruction algorithm based on learned dictionaries \cite{caballero2014dictionary}, \cite{wang2014compressed}, which we abbreviate by TV and DIC, respectively. Note that the methods in \cite{caballero2014dictionary} and \cite{wang2014compressed} further include a total variation penalty term in the formulation of the reconstruction problem which was reported to further increase the image quality of the reconstruction. However, in order to better compare the effect of the differently learned components of the reconstruction algorithms, we neglect the TV-penalty term for the dictionary learning-based reconstruction. Figure \ref{quantitative_comparisons} shows an example of results obtained by the three different methods.

\subsubsection{Total Variation-Minimization} 
Our first method of comparison is the well known TV-minimization-based reconstruction, see for example \cite{block2007undersampled} or \cite{wang2004image} and \cite{caballero2014dictionary}, in the case the dictionary learning-based regularization term is neglected. 
The reconstruction problem is formulated as 
\begin{equation}\label{TV_min_problem}
\underset{\xx}{\mathrm{min}} \, \frac{1}{2}\| \Au \xx  - \yu \|_2^2 + \frac{\lambda}{2} \| \mathbf{G}\xx \|_1,
\end{equation}
where $\mathbf{G}$ denotes the discretized version of the isotropic first order finite differences filter in all three dimensions. Problem (\ref{TV_min_problem}) is solved by ADMM as in \cite{caballero2014dictionary}, by introducing an auxiliary variable $\mathbf{z}$ and alternatively solving for $\mathbf{z}$ and $\xx$. For updating $\mathbf{z}$, an iterative shrinkage method is used, see \cite{chambolle2005total}. Updating $\xx$ corresponds to solving a problem which is linear in $\xx$ and therefore to solving a system of linear equations, for which we used the pre-conditioned conjugate gradient method (PCG). We used a total number of $n_{\mathrm{iter}}=16$ iterations for ADMM, where $\zz$ is updated using one iteration of the iterative shrinkage method and the linear system for the second sub-problem for updating $\xx$ is solved by $n_{\mathrm{iter}}=4$ iterations of PCG.

\subsubsection{Dictionary Learning-based Regularization} 
The DIC method for comparison is given by the iterative reconstruction scheme using spatio-temporal learned dictionaries as regularizers presented in \cite{caballero2014dictionary}, \cite{wang2014compressed}. We used the method by neglecting the TV-penalty term.
This means the reconstruction problem is formulated as 
\begin{equation}\label{DIC_reco_problem}
\underset{\xx, \mathbf{D},\{\boldsymbol{\gamma}_j\}_j }{\mathrm{min}}  \| \Au \xx  - \yu \|_2^2 + \frac{\lambda}{2} \sum_{j=1}^p \| \Ed_j (\xx) - \mathbf{D} \boldsymbol{\gamma}_j\|_2^2,
\end{equation}
where $\Ed_j$ is again a patch-extraction operator, $\mathbf{D}$ is a dictionary and $\{\boldsymbol{\gamma}_j\}_j$ is a family of sparse codes. Problem (\ref{DIC_reco_problem}) is also solved via ADMM by alternating the update with respect to $\xx$ and the dictionary $\mathbf{D}$ as well as the sparse codes $\boldsymbol{\gamma}_j$. For learning the dictionary from the current estimate of the solution $\xx$, we performed 10 iterations of the iterative thresholding an $K$ residual means method\cite{schnass2018convergence}, which is a faster alternative to $K$-SVD \cite{aharon2006k}, while for obtaining the sparse codes, we used orthogonal matching pursuit \cite{tropp2007signal}. As in \cite{caballero2014dictionary} and \cite{wang2014compressed}, the dictionary was trained on patches of shape $4 \times 4 \times 4.$ However, in contrast to the original works, we found a sparsity level $S=16$ and a number of atoms of $K=d=64= 4 \cdot 4 \cdot 4$ to deliver more accurate results. This can most probably be related to the fact that, in contrast to \cite{wang2014compressed} and \cite{caballero2014dictionary}, our $k$-space data is acquired along radial trajectories and the undersampling artefacts have an inherently different structure from the ones obtained by Cartesian sampling.

\subsection{Results}

Table \ref{quantitative_comparisons_table} summarizes the results obtained by the two just introduced methods of comparison and our proposed approach. The Table was obtained by averaging the measures over all $N_z=36$ slices of the all four patients which we reconstructed for all methods. 
The first column shows the measures corresponding to the NUFFT-reconstruction which is directly obtained from the measured $k$-space data. The second column shows the results obtained by the TV-minimization approach which increases image quality with respect to all reported measures. The DIC method and our approach further improve the image quality as can be seen by a further increase of SSIM, HPSI and PSNR and decrease of NRMSE. However, our proposed reconstruction considerably outperforms DIC by $\approx 5.3$ dB in terms of PSNR, $\approx 7\%$ in terms of SSIM and $0.024$ in terms of NRMSE. The increase of HPSI on the other hand, is relatively small.\\
\begin{figure}
\resizebox{\linewidth}{!}{
\begin{overpic}[height=6.9cm,tics=10]{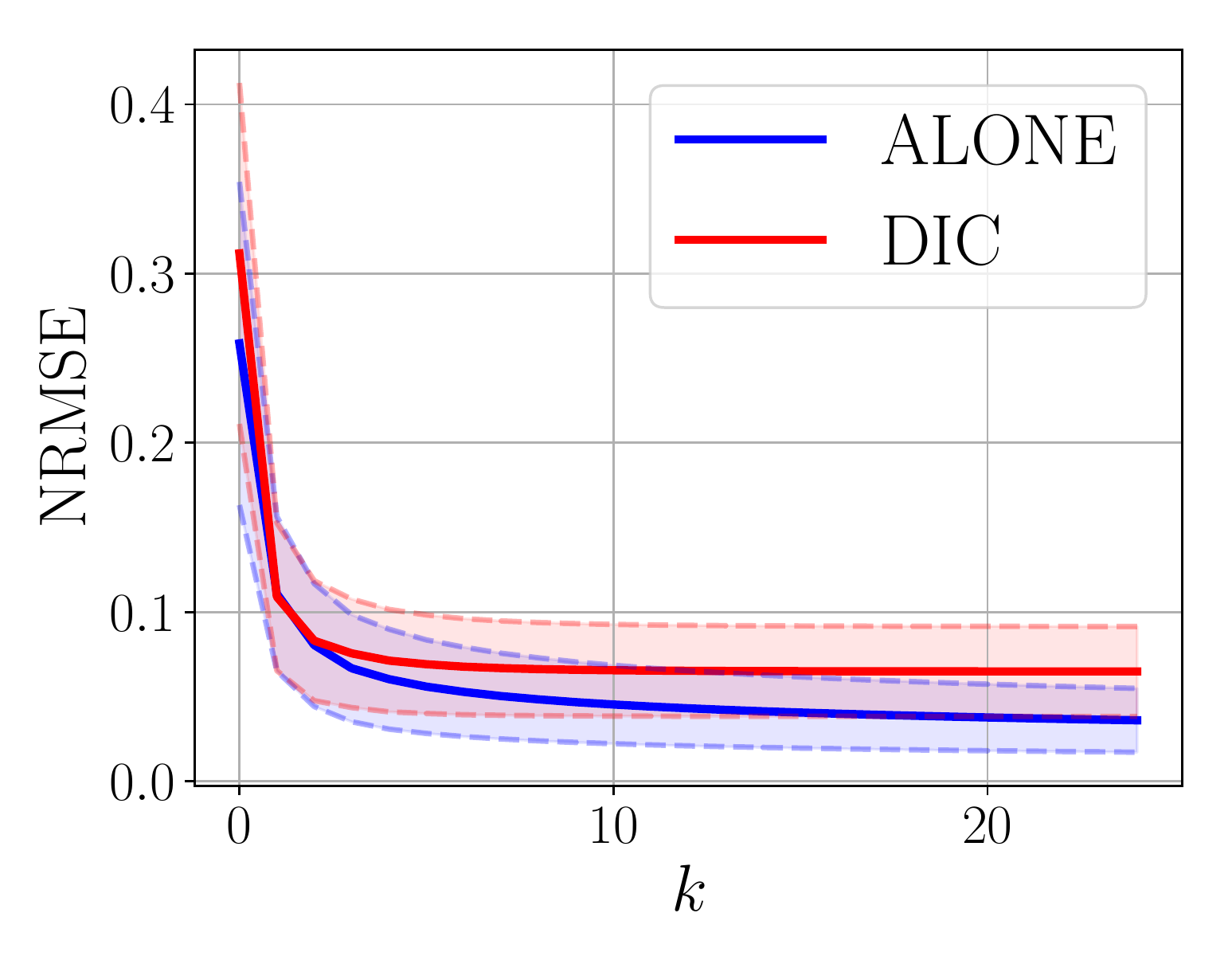}
 \put (76,78) {\small\textcolor{black}{(a)}}
\end{overpic}\hspace{-0.1cm}
\begin{overpic}[height=6.9cm,tics=10]{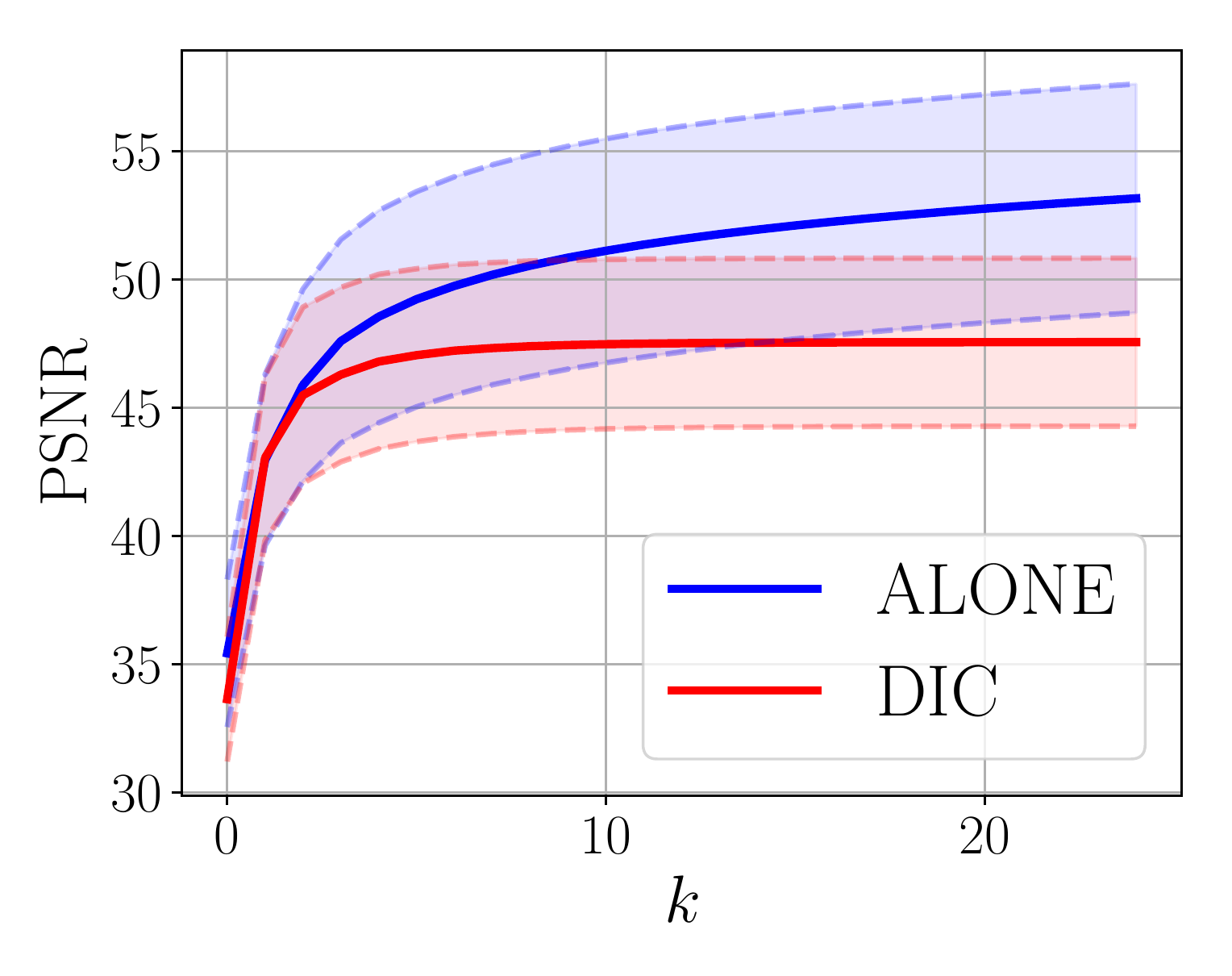}
 \put (76,78) {\small\textcolor{black}{(b)}}
\end{overpic}\hspace{-0.1cm}
}
\resizebox{\linewidth}{!}{
\begin{overpic}[height=6.9cm,tics=10]{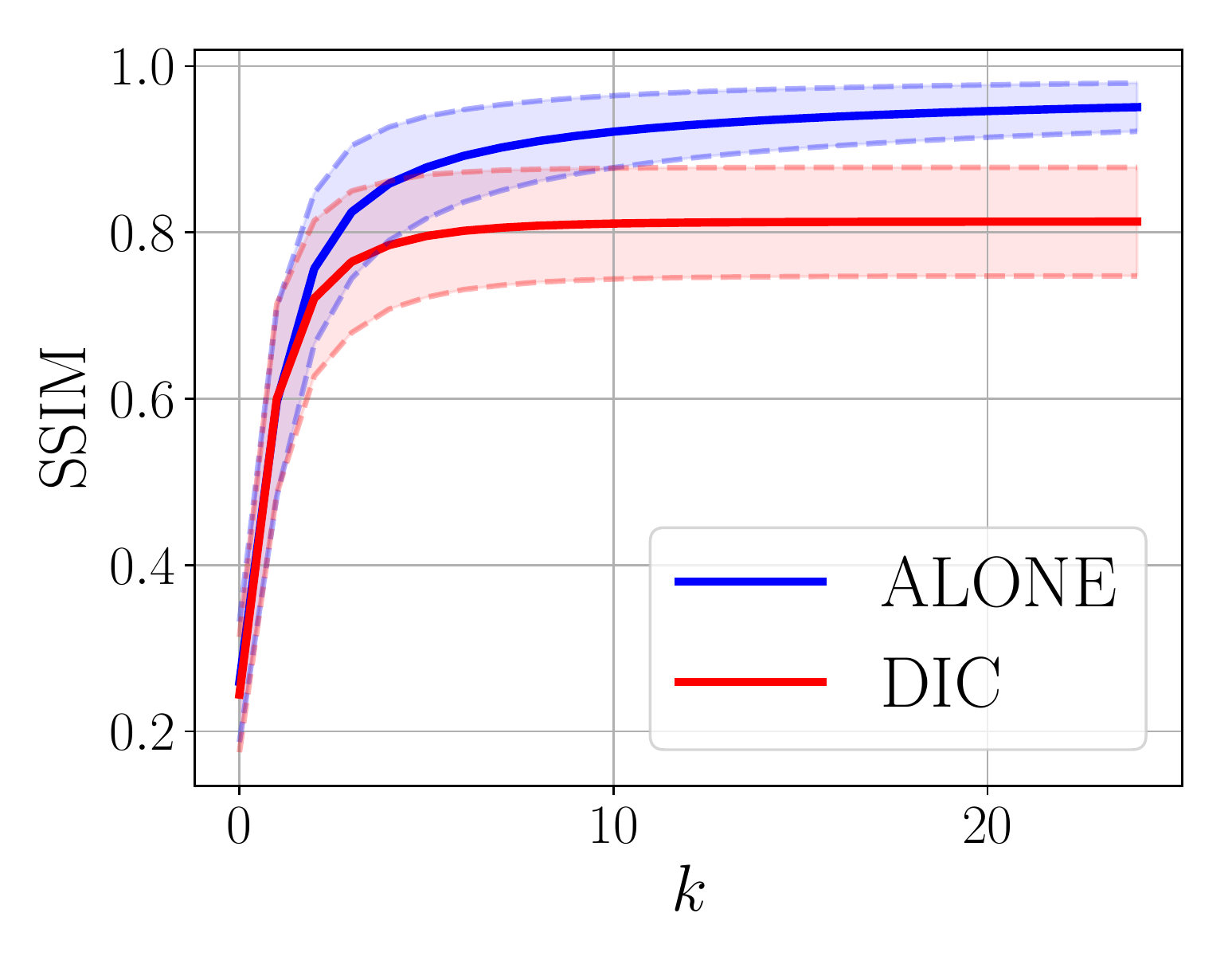}
 \put (76,78) {\small\textcolor{black}{(c)}}
\end{overpic}\hspace{-0.1cm}
\begin{overpic}[height=6.9cm,tics=10]{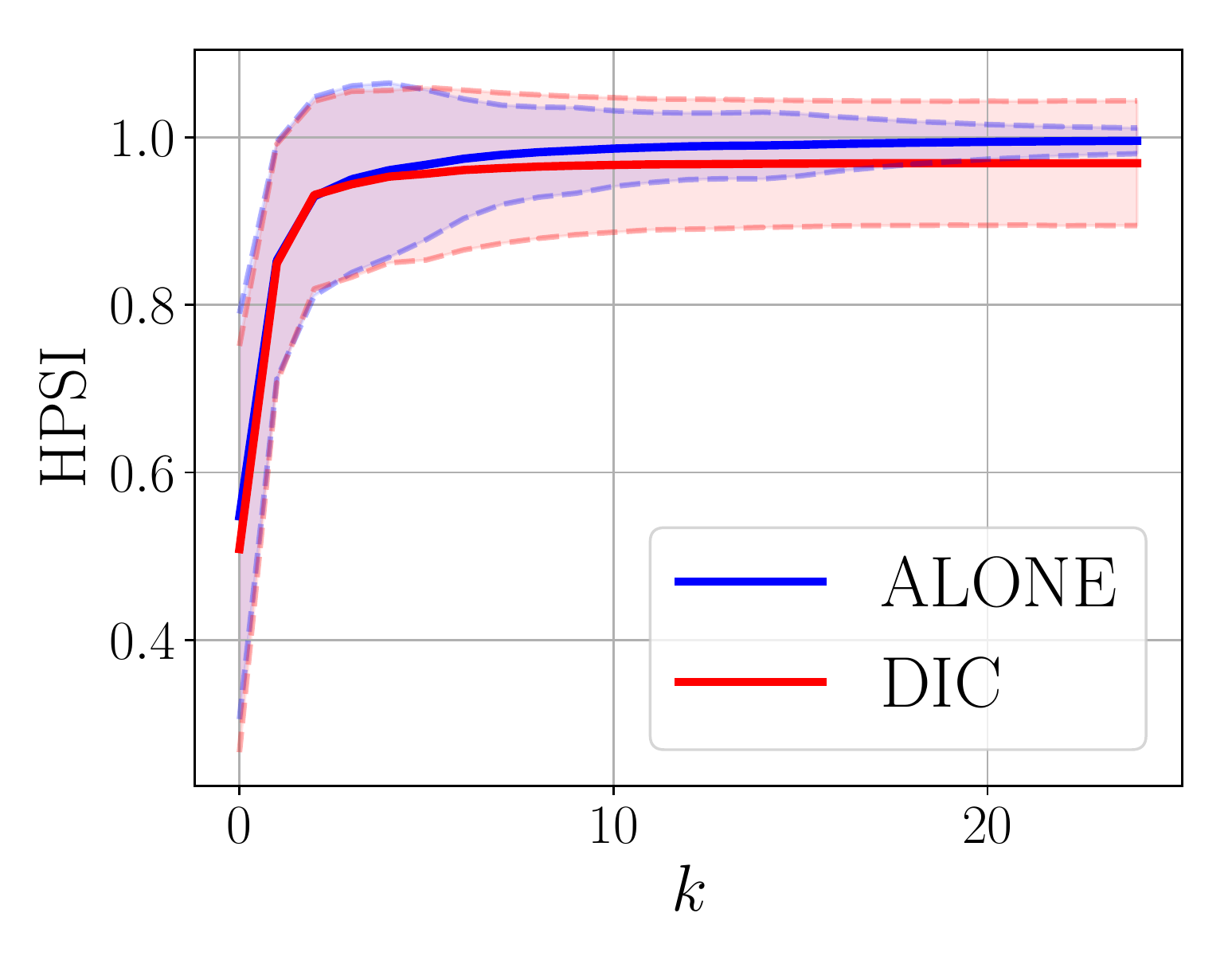}
 \put (76,78) {\small\textcolor{black}{(d)}}
\end{overpic}\hspace{-0.1cm}
}
\caption{Convergence behaviour of the different learning-based methods with respect to the reported quantitative measures. The solid lines correspond to the mean value of the statistics calculated over the complete dataset. The dashed lines denote given by the average measured $\pm$ the corresponding standard deviation.}\label{convergence_behaviour}
\end{figure}
Since our proposed method and the DIC method approach are similar in the sense that the regularization is adaptively learned from the current image estimate during the iterative reconstruction, we investigated the convergence behaviour of the two methods. Figure \ref{convergence_behaviour} shows different curves for the different reported quantitative measures during the reconstruction. The solid lines correspond to the mean value of the measure averaged over the complete dataset of $N_z=36$ slices. Further, the dashed lines show the curves given by the mean $\pm$ the standard deviation of the considered measure. The measures were calculated after having solved the system $\Hd \xx = \cc_k$, where $\cc_k$ is given as in (\ref{ck}) for our method. For the DIC method, $\cc_k$ is given as in (\ref{ck}) but with $\zz_j$ given as the sparse approximation of all patches of the current image estimate, i.e. $\zz_j = \mathbf{D} \gamma_j$ for all $j$.
Note that, while the DIC methd reaches a point of stagnation in terms of NRMSE decrease (which naturally corresponds the measure which is minimized during the iterative reconstruction) and PSNR between the fifth and tenth iteration, our proposed method ALONE seems to still have the potential to further improve image image quality, as neither NRMSE nor PSNR or SSIM have reached a point of saturation. Further, note how for all measures except for PSNR the standard deviation of the measure becomes smaller during the reconstruction which indicates an improved stability of the algorithm compared to the DIC method.

\begin{table}[!h]
\centering
\renewcommand{\arraystretch}{1.3}
\small{
 \caption{Comparison ALONE with different iterative methods using different regularizations.}\label{quantitative_comparisons_table}
 \sisetup{table-number-alignment = center,table-figures-integer  = 2,table-figures-decimal  = 3,round-mode  = places,round-precision =3}
\centering
\begin{tabular}{ l|
				S[table-column-width = 0cm]
				S[table-column-width = 0cm]
				S[table-column-width = 0cm]
				S[table-column-width = 0cm]}
\toprule
   &  \, \textbf{NUFFT}  &  \, \, \textbf{TV} & \, \, \textbf{DIC }  &  \textbf{ALONE}  \\
   \midrule       
   \textbf{PSNR}   & 35.4959 & 40.4122 & 42.8583 &  48.1221 \\
     \textbf{SSIM} & 0.6258 & 0.8378 & 0.8949 &  0.9616\\
     \textbf{HPSI} & 0.9551 & 0.9814 & 0.9907  &   0.9977 \\
    \textbf{NRMSE} & 0.1371 & 0.0770 & 0.0577 &  0.0334 \\
    \bottomrule
  \end{tabular}
  }
\end{table}

\section{Discussion}\label{section_discussion}

In this work, we have presented a simple yet powerful method named ALONE (Adaptive Learning of NEtworks) for regularization for 2D undersampled radial cine MRI. The method is based on the adaptive regularization of the solution given in the form of a shallow CNN which is trained in an unsupervised manner during the reconstruction. \\
We have compared ALONE to a well-known total variation-minimization approach (TV) as well as to another learning-based method which employs an adaptive regularization based on dictionary learning (DIC). Our method outperforms the TV- and the DIC method with respect to all reported measures. Further, we investigated the effect of the adaptive regularization for both learning-based methods during the reconstruction.  ALONE shows an improved and more stable convergence behaviour during the reconstruction which is visible in terms of a smaller standard deviation of NRMSE, SSIM and HPSI during the reconstruction.\\
Further, our proposed approach ALONE has one significant advantage over the dictionary learning-based method DIC, which is the acceleration of the regularization step during the reconstruction. Note from (\ref{DIC_reco_problem}), that the reconstruction problem is formulated as joint minimization problem over the variables $\xx$, $\mathbf{D}$ and $\{\boldsymbol{\gamma}_j\}$. In contrast, from \eqref{eq:main} we see that our formulation only requires the update of two variables. While for DIC, training the dictionary $\mathbf{D}$ is achieved in a relatively short time, the computational bottleneck of the approach is  finding the sparse codes $\boldsymbol{\gamma}_j$ of the patches $\Ed_j (\xx)$ with respect to the dictionary $\mathbf{D}$. This is because obtaining $\boldsymbol{\gamma}_j$ involves solving  an optimization problem for all $j$, namely the sparse coding problem. On the other hand, the sparse-approximation counterpart in our reconstruction scheme is given by calculating $\zz_{j} = \net_\theta\big(\Ed_j (\xx)\big) $, i.e.\ by performing a forward pass of the patches through the (shallow) CNN. Table \ref{reco_times_table} shows a direct comparison of the corresponding counterparts for the DIC method and our proposed reconstruction scheme ALONE. The time in the Table refers to the average time needed for obtaining the regularized image, i.e. for the patch-wise sparse approximation using OMP for the DIC method, and for obtaining $\cc_k$ for our method ALONE by performing a forward pass of all patches. Therefore, the total cost of the regularization can be estimated by multiplying the average time by the number of iterations $T$ one sets before the reconstruction.

\begin{table}[!h]
\centering
\renewcommand{\arraystretch}{1.3}
\small{
 \caption{Comparison of the different components for the regularization with dictionary learning and our proposed method ALONE. }\label{reco_times_table}
 \sisetup{table-number-alignment = center,table-figures-integer  = 2,table-figures-decimal  = 3,round-mode  = places,round-precision =3}
\centering
\begin{tabular}{ l | c	c }
\toprule
    &   \textbf{DIC}  & \textbf{ALONE}  \\
\midrule
Patches size  &   $4 \times 4 \times 4$  &  $32 \times 32 \times 4$ \\
Strides & $2 \times 2 \times 2$ &  $16 \times 16 \times 2$ \\ 
Number of patches & 353\,934 & 5\,054\\
\midrule
Training & ITKrM & Back-propagation \\
Time & $\approx$ 10 s &  $\approx$ 6 s\\
\midrule
Patches approximation & OMP & Forward pass \\
Time & $\approx 7$ m & $\approx 0.3$ s \\
\midrule
Data type of patches & $\mathbb{R}$ & $\mathbb{C}$ \\ 
 \bottomrule
  \end{tabular}
  }
\end{table}

\begin{figure}[!h]			
\begin{minipage}{\linewidth}
\centering
\resizebox{\linewidth}{!}{
\includegraphics[height=1.9cm]{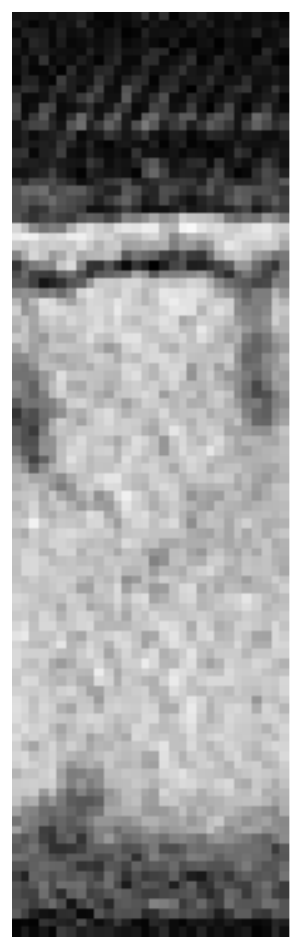}\hspace{-0.2cm}
\begin{overpic}[height=1.9cm,tics=10]{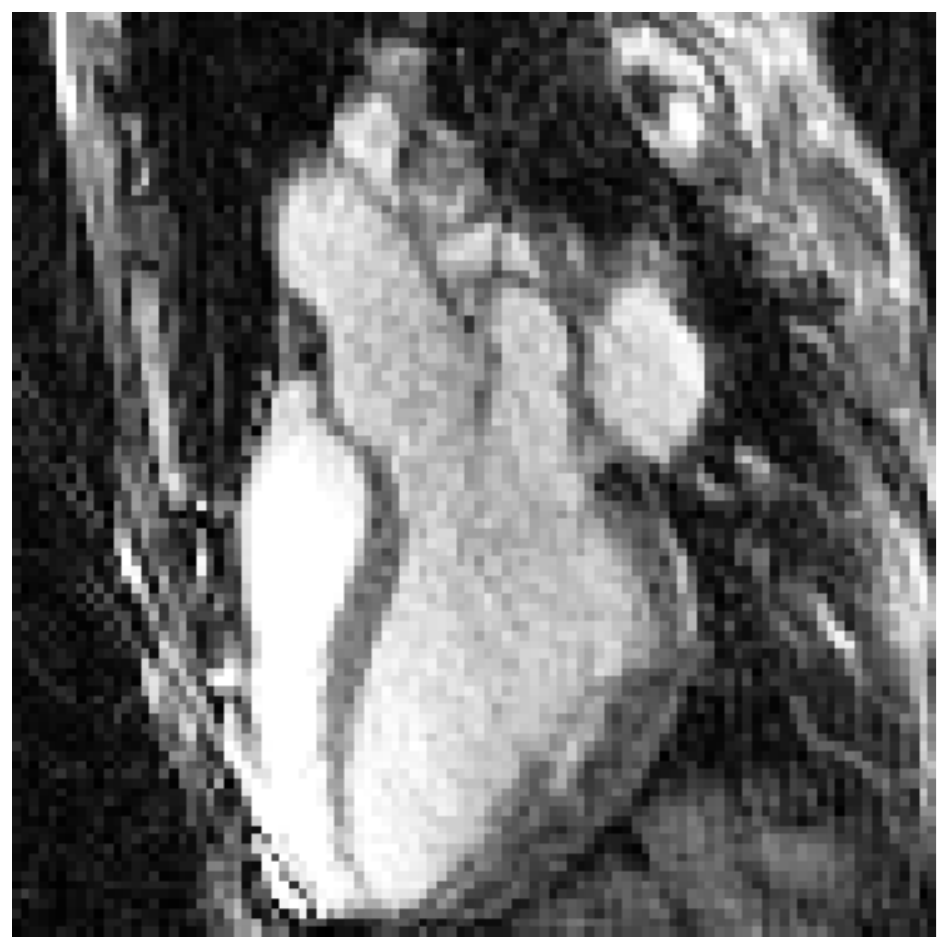}
 \put (76,8) {\small\textcolor{white}{(a)}}
\end{overpic}\hspace{-0.1cm}
\includegraphics[height=1.9cm]{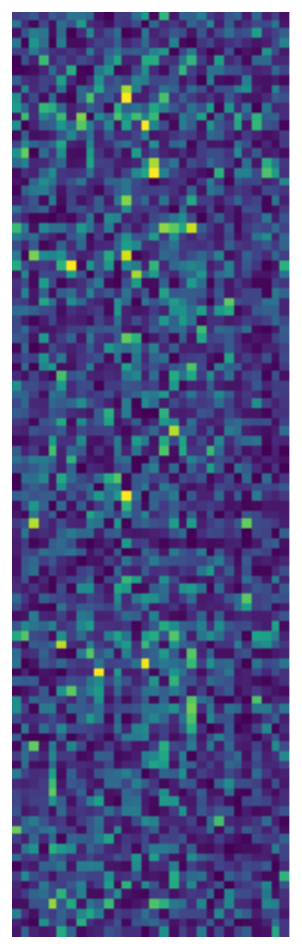}\hspace{-0.2cm}
\begin{overpic}[height=1.9cm,tics=10]{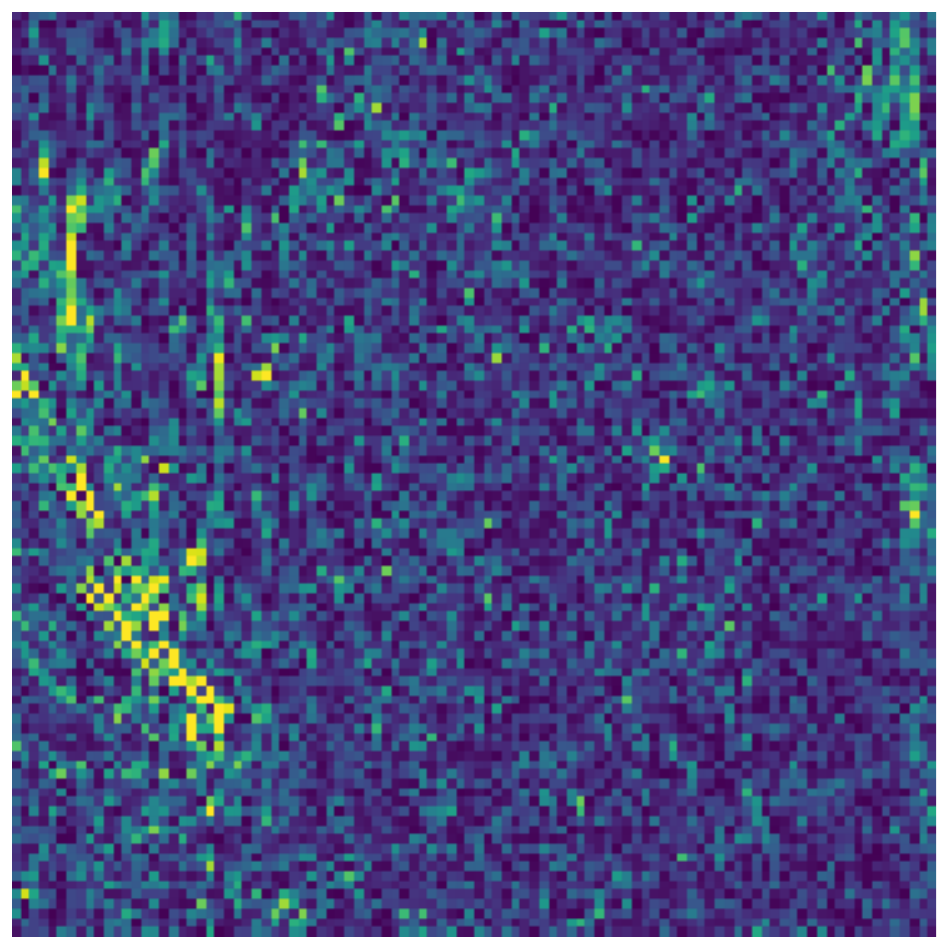}
\end{overpic}
}\\
\resizebox{\linewidth}{!}{
\includegraphics[height=1.9cm]{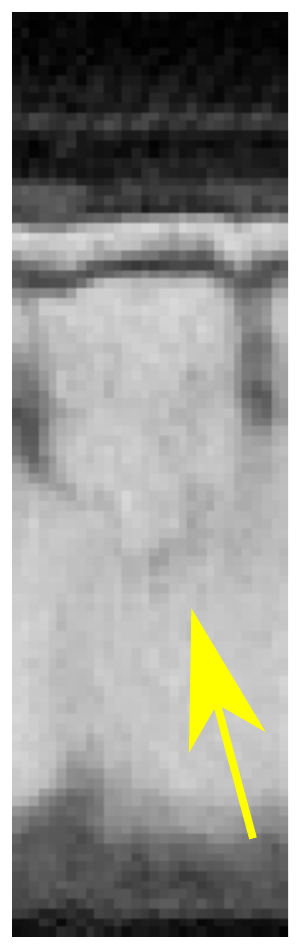}\hspace{-0.2cm}
\begin{overpic}[height=1.9cm,tics=10]{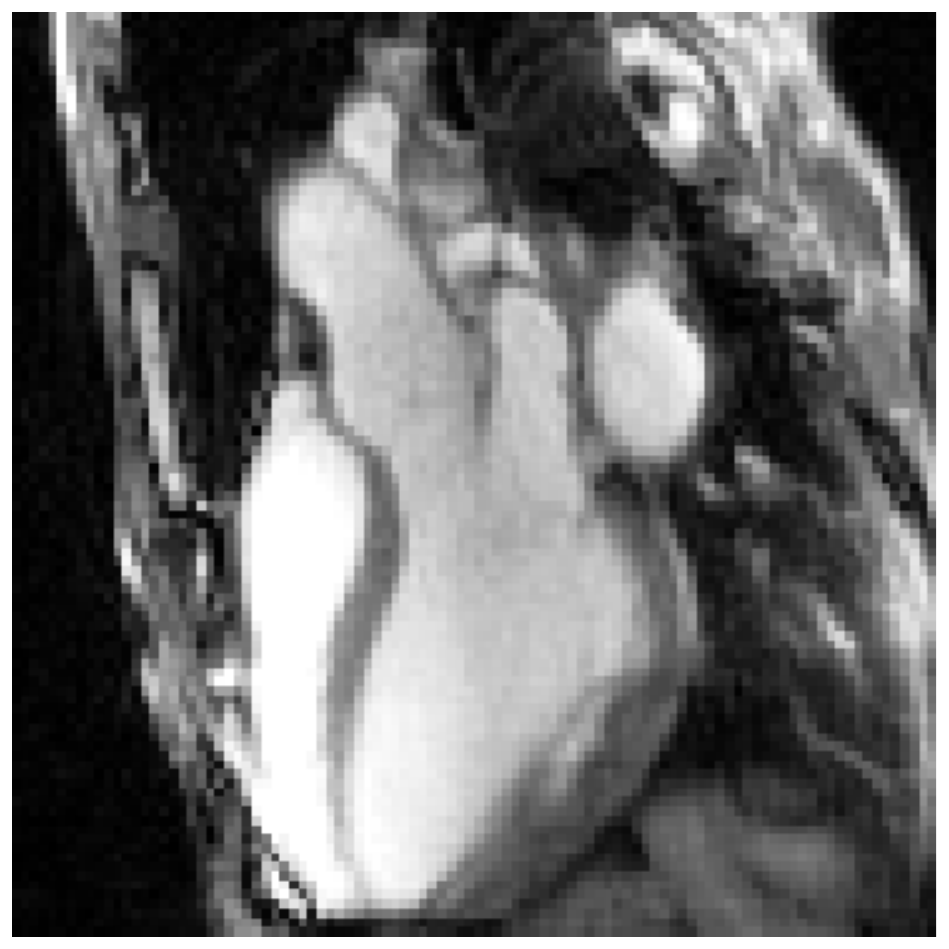}
 \put (76,8) {\small\textcolor{white}{(b)}}
\end{overpic}\hspace{-0.1cm}
\includegraphics[height=1.9cm]{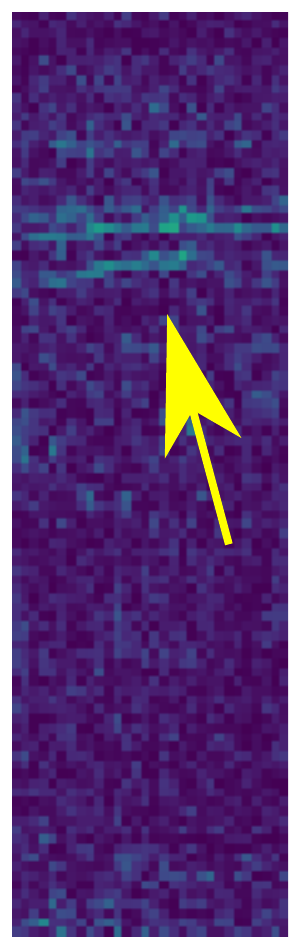}\hspace{-0.2cm}
\begin{overpic}[height=1.9cm,tics=10]{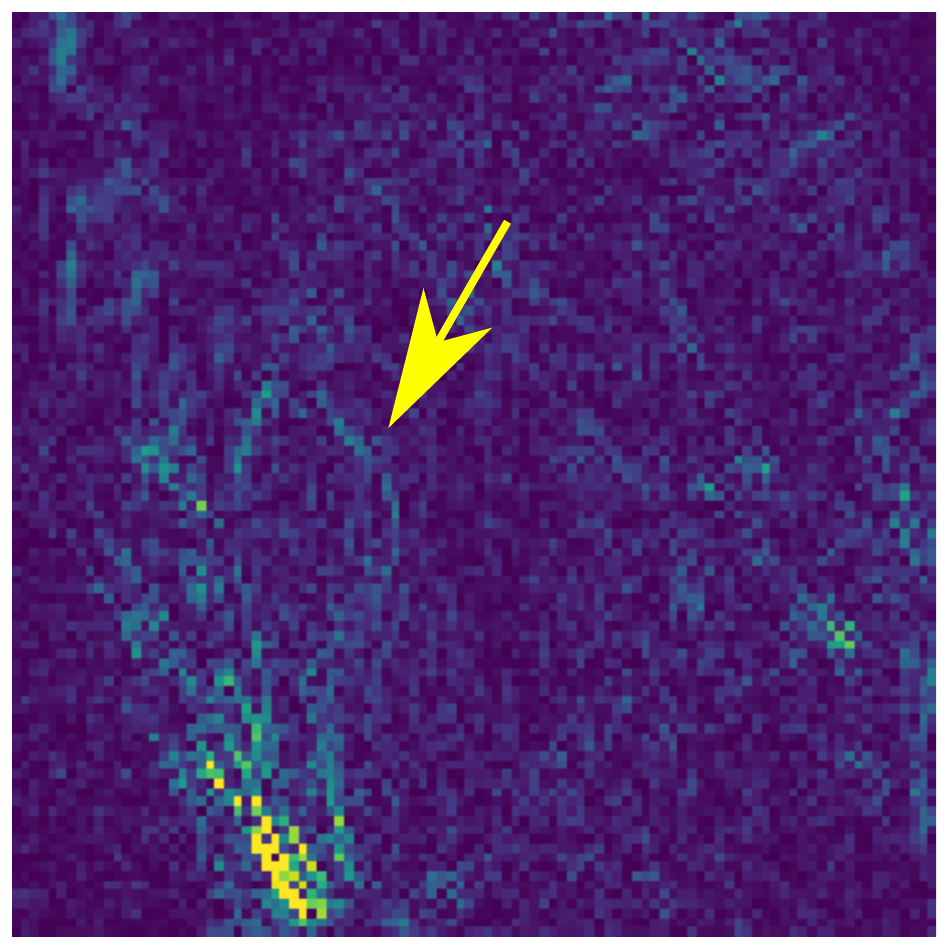}
\end{overpic}
}\\
\resizebox{\linewidth}{!}{
\includegraphics[height=1.9cm]{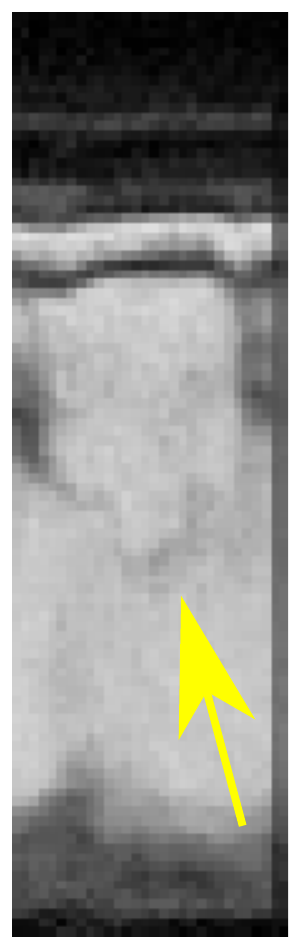}\hspace{-0.2cm}
\begin{overpic}[height=1.9cm,tics=10]{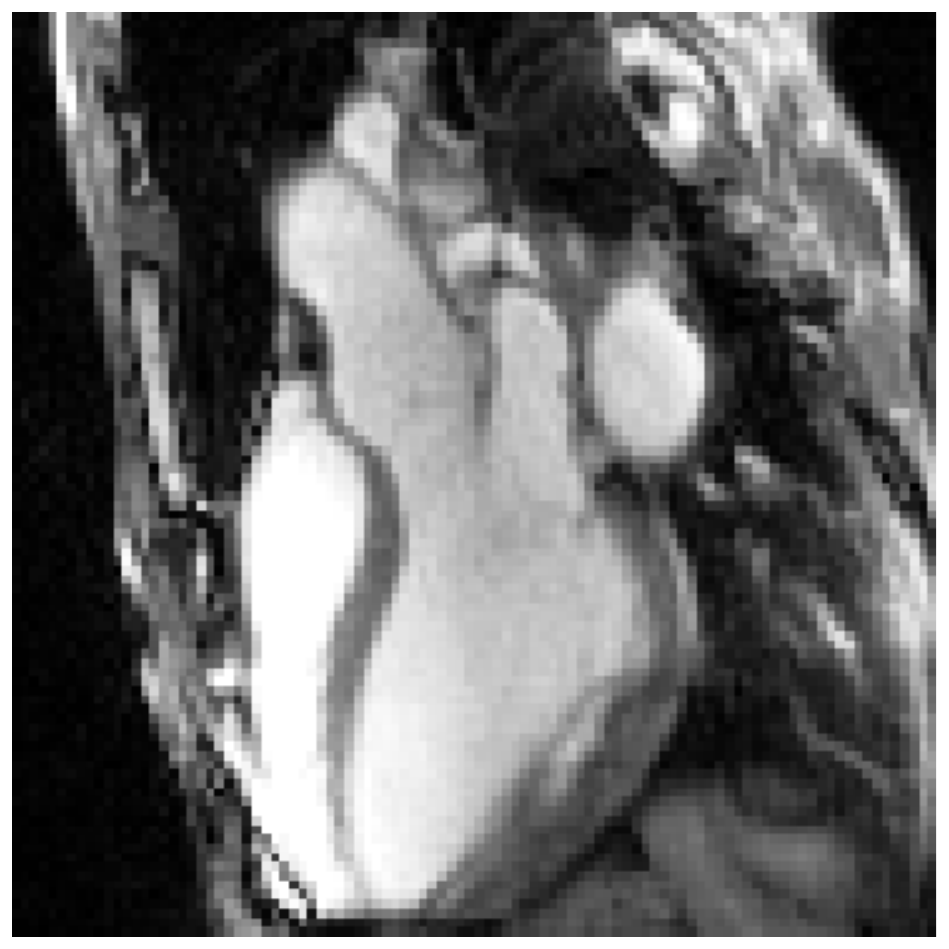}
 \put (76,8) {\small\textcolor{white}{(c)}}
\end{overpic}\hspace{-0.1cm}
\includegraphics[height=1.9cm]{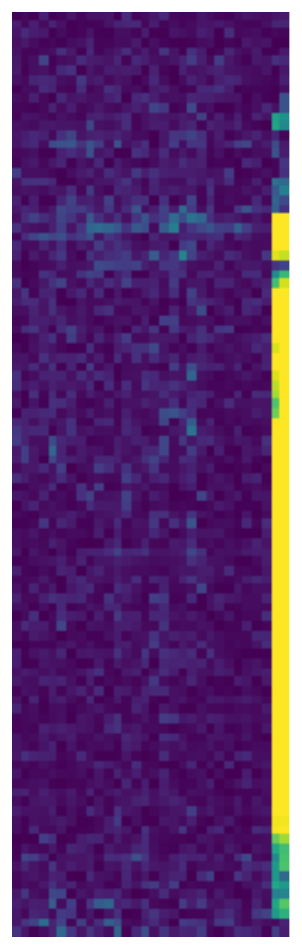}\hspace{-0.2cm}
\begin{overpic}[height=1.9cm,tics=10]{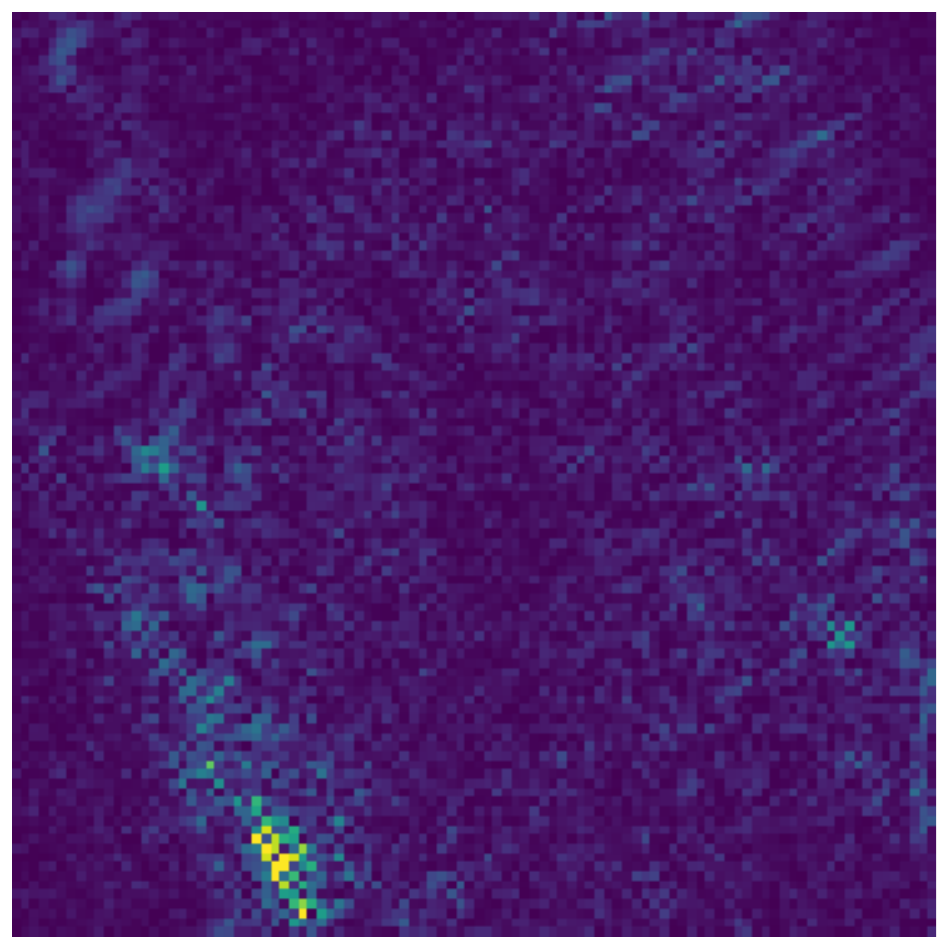}
\end{overpic}
}\\
\resizebox{\linewidth}{!}{
\includegraphics[height=1.9cm]{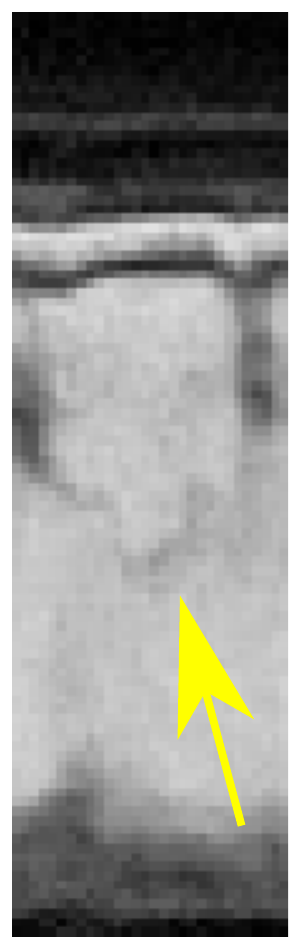}\hspace{-0.2cm}
\begin{overpic}[height=1.9cm,tics=10]{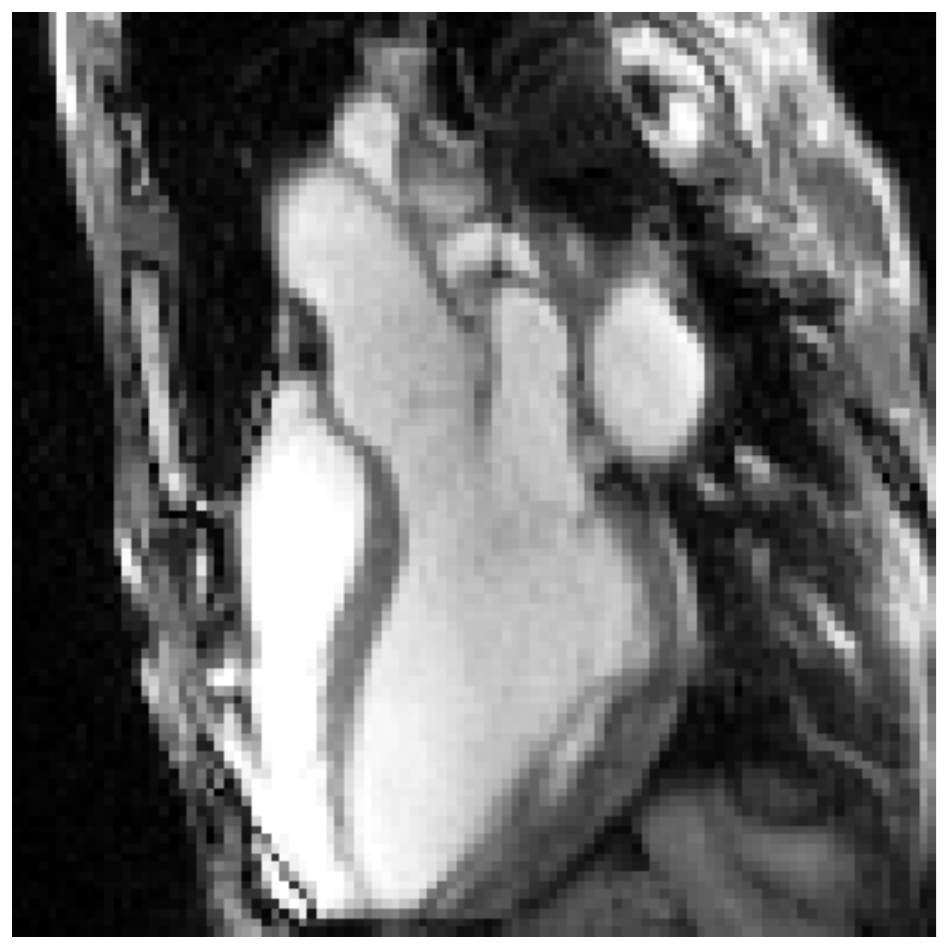}
 \put (76,8) {\small\textcolor{white}{(d)}}
\end{overpic}\hspace{-0.1cm}
\includegraphics[height=1.9cm]{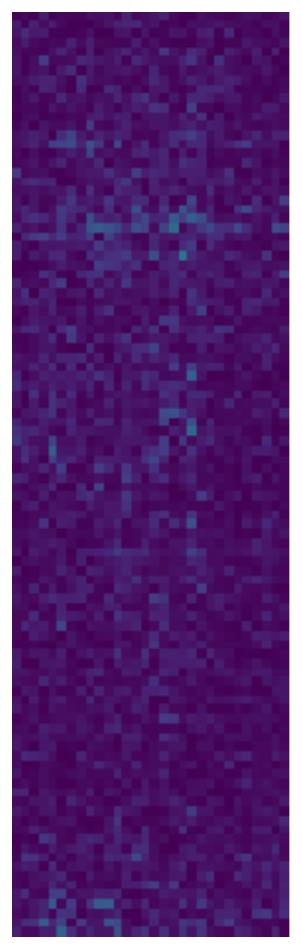}\hspace{-0.2cm}
\begin{overpic}[height=1.9cm,tics=10]{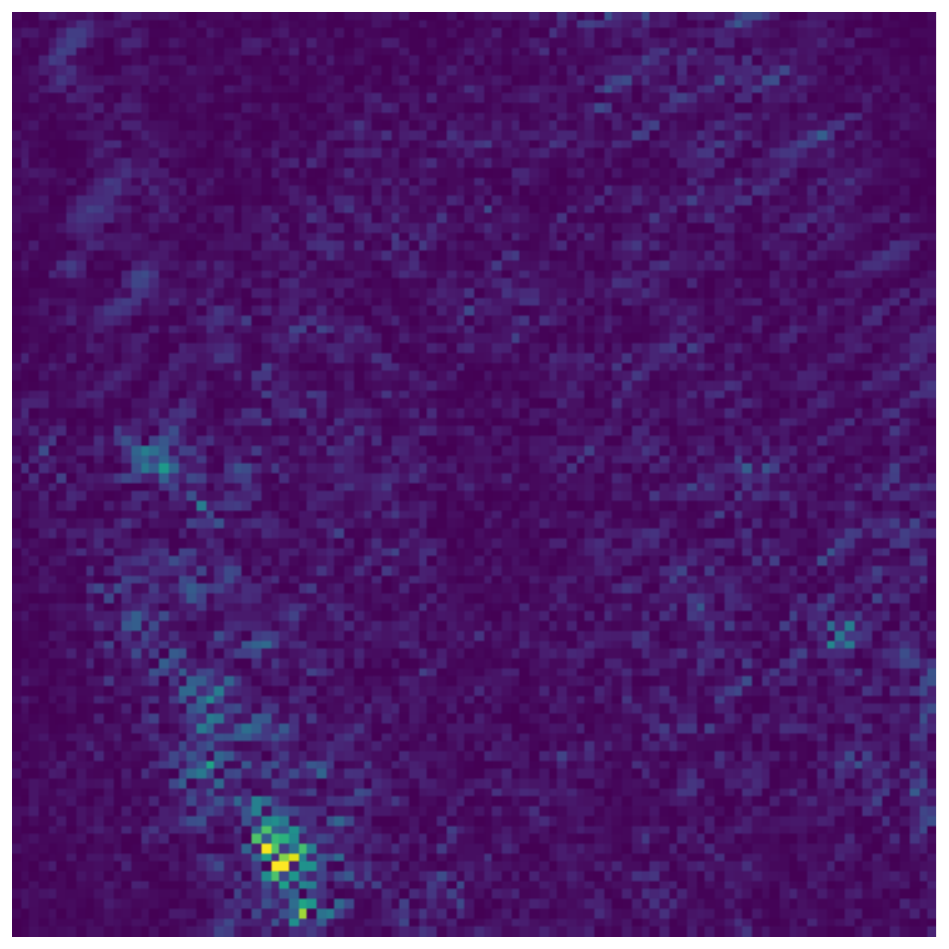}
\end{overpic}
}\\
\resizebox{\linewidth}{!}{
\includegraphics[height=1.9cm]{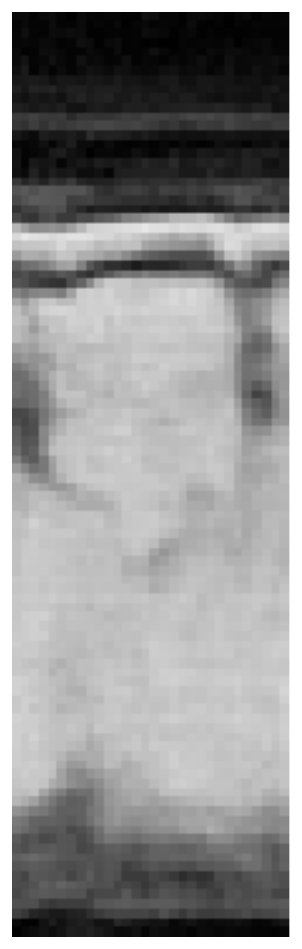}\hspace{-0.2cm}
\begin{overpic}[height=1.9cm,tics=10]{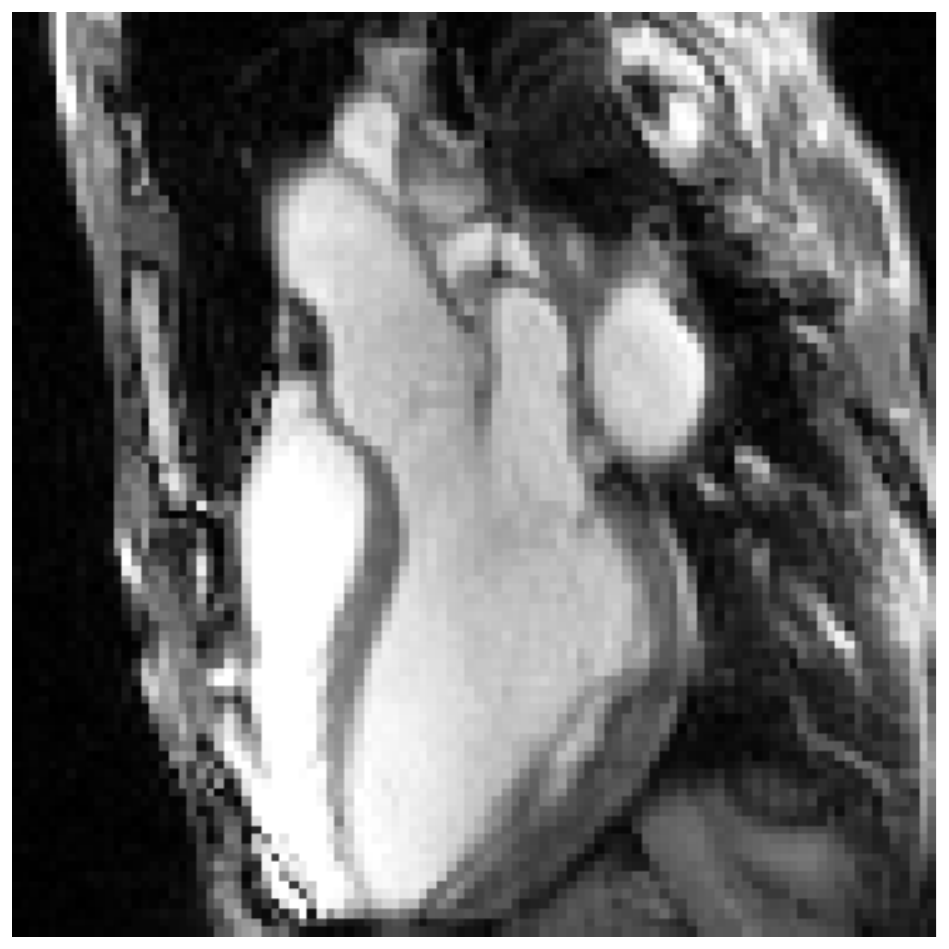}
 \put (76,8) {\small\textcolor{white}{(e)}}
\end{overpic}\hspace{-0.1cm}
\includegraphics[height=1.9cm]{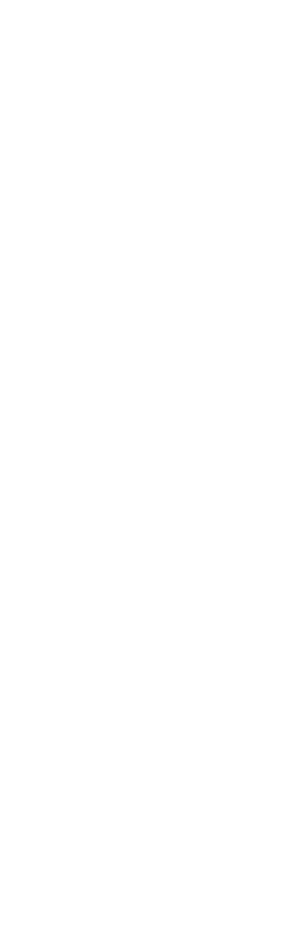}\hspace{-0.2cm}
\begin{overpic}[height=1.9cm,tics=10]{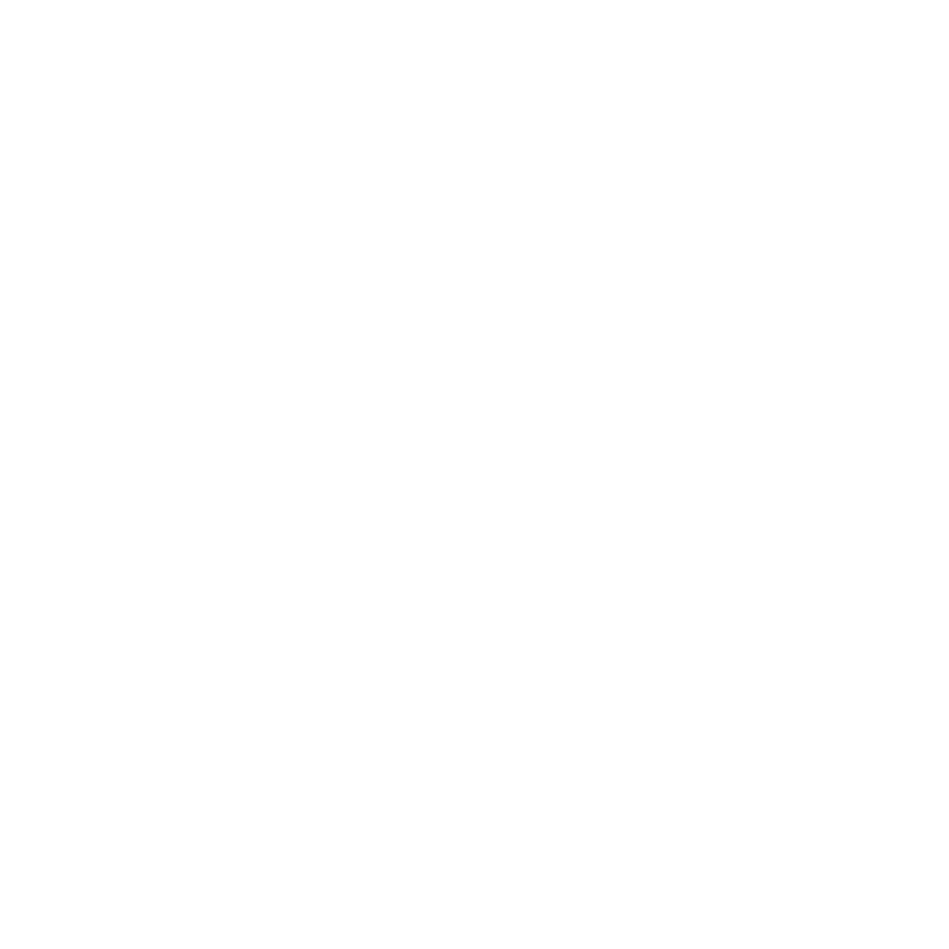}
 \put (76,8) {\small\textcolor{white}{(f)}}
\end{overpic}
}
\\
\end{minipage}\\
\caption{Results obtained by the NUFFT reconstruction from $N_\varphi=1130$ radial spokes (a), with the DIC method (b), with ALONE by using a complex-valued CNN $\net_\theta$ (c), with ALONE by using a real-valued CNN $\net_\theta$ (d) and the reference $kt$-SENSE reconstruction obtained from $N_{\varphi} = 3400$ radial spokes.} \label{real_vs_complex_fig}
\end{figure}

In \cite{caballero2014dictionary}, it was reported that for the DIC method, using real-valued dictionaries for the sparse approximation of the complex-valued images outperformed the usage of complex-valued dictionaries. Note that this adds another (non-negligible) factor of two to the most computational demanding component of the reconstruction, i.e. the sparse coding of all patches. In our proposed reconstruction scheme ALONE, in contrast, there is no noticeable difference between using a real-valued and a complex-valued CNN in terms of speed, as the additional increase of complexity is negligible. For ALONE, training the CNN on complex-valued patches represented by two input-channels yielded more accurate reconstruction. 
Since iterative reconstruction is time consuming, we only reconstructed the image data for only one of the four patients with $N_z=12$ slices by employing a CNN-based regularization which is learned from the real-valued patches. Similarly to \cite{caballero2014dictionary}, the complex-valued patches were then obtained by performing a forward pass of the real and the imaginary part of the patches using the same CNN. Figure \ref{real_vs_complex_fig} shows an example of results obtained with ALONE using a real-valued and complex-valued CNN, where we see that the  complex-valued CNN improved the results. Further, from the point-wise error images and the yellow arrows, we see that DIC  tends to slightly smooth image details, while ALONE well preserves edges. Table \ref{real_vs_complex_table} shows the obtained results for one of the patients for the real- and the complex version of our proposed ALONE reconstruction as well as for the DIC method.\\

\begin{table}[!h]
\centering
\renewcommand{\arraystretch}{1.3}
\small{
 \caption{Comparison of ALONE using a real-valued and complex-valued CNN $\net_\theta$ to DIC using real valued dictionaries.}\label{real_vs_complex_table}
 \sisetup{table-number-alignment = center,table-figures-integer  = 2,table-figures-decimal  = 3,round-mode  = places,round-precision =3}
\centering
\begin{tabular}{ l|
				S[table-column-width = 0cm]
				S[table-column-width = 0cm]
				S[table-column-width = 0cm]
				S[table-column-width = 0cm]}
\toprule
   &  \textbf{DIC} ($\mathbb{R}$) & \textbf{ALONE} ($\mathbb{R}$)  &   \textbf{ALONE} ($\mathbb{C}$)  \\
   \midrule       
   \textbf{PSNR}  &  40.6229 & 43.6687 & 45.5045\\ 
     \textbf{SSIM} & 0.8675 & 0.9262 & 0.9498 \\
     \textbf{HPSI} & 0.9937 & 0.9976 & 0.9985 \\
    \textbf{NRMSE} &  0.0589 & 0.0426  & 0.0352 \\
    \bottomrule
  \end{tabular}
  }
\end{table}

Another advantage of using ALONE is that we can use relatively large patch-sizes and strides without observing artefacts which could be attributed to the overlapping of patches. The reason for that most probably lies in the fact that the network $\net_\theta$ is a  CNN. Since CNNs are well-known to be translation-equivariant, using larger strides for the regularization of  the solution does not lead to block-artefacts. In contrast, since the approximation using a learned dictionary can be identified with a single fully-connected layer taking the sparse code $\boldsymbol{\gamma}_j$ as an input, the operation is not translation-equivariant in general and therefore, relatively small strides need to be used in order to avoid patchy artefacts. Compared to the DIC method, in our proposed method ALONE, the number of patches needed to be processed is lower, larger strides can be used and the patch-wise approximation is faster. In sum, this results in an acceleration of the regularization step by several orders of magnitude. Therefore, for ALONE, the overall cost of the reconstruction is dominated by the application of the forward and adjoint operators as in any other reconstruction algorithm.\\
Even if the regularization of the solution in each iteration is highly accelerated, the main limitation of the method clearly remains the relatively high number of iterations needed to perform the reconstruction. Further, the strength of the regularization $\lambda$ has to be chosen a priori which might be problem-dependent.
However, note that the usage of cascaded networks, which can be thought of unrolled iterative schemes is prohibitive for large-scale problems as the one considered in this work. Opposed to other works using cascaded networks for the MR image reconstruction task, see e.g. \cite{schlemper2017deep}, \cite{hammernik2018learning}, \cite{qin2018convolutional}, our forward operator $\Au$ is given by a radial Fourier encoding operator with $n_c=12$ coils and does not allow the construction of iterative neural networks due to its computational complexity. For a more detailed discussion about the issue and a possible way to overcome the problem, we refer to \cite{kofler2019}.\\

\section{Conclusion}\label{section_conclusion}

In this work we have presented a new reconstruction algorithm named for accelerated 2D radial cine MRI. The reconstruction algorithm involves a patch-wise regularization which is adaptively learned during the reconstruction in an unsupervised manner. Therefore, the method does not require having access to large training datasets with ground truth data. \\
We have compared our reconstruction method to a total variation-minimization method and to a dictionary learning-based method using adaptively trained dictionaries. Our method outperformed both methods with respect to all chosen reported measures. Further, compared to the dictionary learning-based method, it accelerates the reconstruction by orders of magnitude since it highly reduces the regularization step needed during the reconstruction. \\ 
While in this work we have applied our reconstruction method ALONE to 2D cardiac radial cine MRI, the method's formulation is held general and therefore we expect ALONE to be applicable to other imaging modalities as well.

\section*{Acknowledgements}
A. Kofler and M. Dewey acknowledge the support of the German Research Foundation (DFG), project number GRK 2260, BIOQIC.   The work of M. Haltmeier  has been supported by the Austrian Science Fund (FWF), project P 30747-N32.

\appendix[Mathematical Analysis of ALONE]

In this appendix we present some theoretical  results for Algorithm~\ref{reco_algo} (ALONE). For that purpose, we assume absence of noise in the measurements
and note that  ALONE is of the fixed point form
\begin{align*} 
\theta_{k}  
& \in  \argmin_\theta  
\frac{\lambda}{2} \norm{\Ed (\xx_k)- \net_\theta \Ed (\xx_k)}_2^2 + \pen(\theta) \,,
\\   
\xx_{k+1}
&\in  (\Au^\herm\Au + \la \Ed^\trans \Ed )^{-1} (\Au^\herm\yu  + \la \Ed^\trans \net_{\theta_{k}} \Ed \xx_{k} )  \,,
\end{align*}
with initialization  $\xx_0 = \Au^\herm \yu$.

\subsection{Characterization of fixed points}

We first define the natural underlying prior  information for solutions of  \eqref{inv_problem} induced by ALONE. 

\begin{definition}[$\theta^*$-adapted solution] 
For any $\theta^* \in \R^q$, we  call $\xx^* \in \C^N$ a $\theta^*$-adapted 
solution of  \eqref{inv_problem}, if     
\begin{align} \label{eq:ip1}
& \Au \xx^* = \yu 
\\ \label{eq:ip2}
& \Ed \xx^*  =  \net_{\theta^*} \Ed \xx^*
\end{align}
\end{definition}

We will show that $\theta^*$-adapted solutions
are fixed points of the ALONE algorithm as well 
as partial minimizers of  $\reg_{\la, \yu}$ 
defined next.

\begin{definition}[Fixed points]
The pair $(\theta^*, \xx^*) \in \C^N \times \R^q$ is called fixed point of 
ALONE if 
\begin{align} \label{eq:fix1}
\theta^*  
& \in  \argmin_\theta   \frac{\lambda}{2} \norm{\Ed \xx^* - \net_\theta( \Ed \xx^*)}_{\Ed}^2 + \pen(\theta^*)
\\ \label{eq:fix2}
\xx^*  
& \in (\Au^\herm\Au + \la \Ed^\trans \Ed )^{-1} (\Au^\herm\yu  + \la \Ed^\trans \net_{\theta^*} \Ed \xx^*) \,.
\end{align}
\end{definition}

\begin{definition}[Partial minimizers]  
The pair  $ (\xx^*, \theta^*) \in \C^N \times \R^q$ is called a 
partial minimizer of $\reg_{\la, \yu}$ if 
\begin{align}  
\label{eq:opt1}
\forall \xx \in \C^N \colon \quad 
& \reg_{\la, \yu}(\xx^*, \theta^*) \leq  \reg_{\la, \yu}(\xx, \theta^*)
\\  \label{eq:opt2}
\forall \theta \in  \R^q \colon \quad 
&  \reg_{\la, \yu}(\xx^*, \theta^*) \leq  \reg_{\la, \yu}(\xx^*, \theta) 
\end{align}  
\end{definition}

We have the following result relating $\theta^*$-adapted 
solutions of inverse problems of the form given in \eqref{inv_problem} to fixed points of ALONE and partial minimizers of 
$\reg_{\la, \yu}$.

\begin{theorem}[Fixed points and partial minimizers for $\theta^*$-adapted solutions] \label{thm:char}
Let $\theta^* \in \R^q$ satisfy  \eqref{eq:fix1} and $\xx^* \in \C^N$ be a $\theta^*$-adapted solution of \eqref{inv_problem}.  
Then the following hold:
\begin{enumerate}[label=(\alph*)]
\item \label{char1} $(\theta^*, \xx^*)$ is a fixed point of ALONE.
\item  \label{char2}
$(\theta^*, \xx^*)$ is a partial minimizer of $\reg_{\la, \yu}$.  
\end{enumerate}         
\end{theorem}

\begin{IEEEproof}
\ref{char1}  Equation \eqref{eq:fix1} is satisfied  by assumption. Further, \eqref{eq:fix2}  is satisfied if and only  if  the optimality condition 
\begin{equation*}
0 = \la (\Ed^\trans \Ed \xx^* - \Ed^\trans \net_{\theta^*} \Ed \xx^*)  +  \Au^\herm(\Au \xx^* - \yu)
\end{equation*}
holds. According to \eqref{eq:ip1}, \eqref{eq:ip2} this is however the case.
Hence  $(\theta^*, \xx^*)$ is a fixed point of ALONE.

\ref{char2} According to \eqref{eq:ip1}, \eqref{eq:ip2}  we have 
$\reg_{\yu,\la}(\xx^*, \theta^*) = \pen(\theta^*)$. Consequently, for any 
$ \xx \in \C^N$ we have 
\begin{multline*}
\reg_{\yu,\la}(\xx, \theta^*) = \frac{\lambda}{2} \norm{\Ed \xx - \net_{\theta^*} \Ed \xx}_2^2  \\+ \frac{1}{2}\norm{\Au \xx - \yu}_2^2 + \pen(\theta^*)  
\geq \pen(\theta^*)   = \reg_{\yu,\la}(\xx^*, \theta^*) \,.
\end{multline*} 
This shows  \eqref{eq:opt1}. 
Using  \eqref{eq:fix1},  one verifies  
\begin{multline*}
\reg_{\yu,\la}(\xx^*, \theta) = \frac{\lambda}{2} \norm{\Ed \xx^* - \net_{\theta} \Ed \xx^*}_2^2  + \pen(\theta) \\+ \frac{1}{2}\norm{\Au \xx^* - \yu}_2^2   
\geq \reg_{\yu,\la}(\xx^*, \theta^*) \,.
\end{multline*} 
Hence \eqref{eq:opt2} is satisfied for any $ \theta \in \R^q$ and
$(\theta^*, \xx^*)$ is a partial minimizer of $\reg_{\la, \yu}$.   
\end{IEEEproof}

\subsection{Existence and  stability}

Next we show that the ALONE  algorithm~\ref{reco_algo} is well-defined (iterates exist) and  stable with respect to data perturbation. For that purpose we assume that the  following reasonable conditions hold.

\begin{assumption}[Existence and  stability]\mbox{}
\begin{enumerate}[label=(A\arabic*), leftmargin = 2.5em]
\item \label{ass:0} $\Au \colon \C^N \to \C^m$ is a  linear forward operator.

\item \label{ass:1} $\R^q \times \C^N \to \C^N \colon (\theta, \xx) \mapsto   \net_\theta(\xx)$ is  continuous. 

\item \label{ass:2}  $\forall j \in \set{1, \dots, p} \colon$ 
$\Ed_j \colon \C^N \to \C^d$ is linear.   

\item \label{ass:3} $\pen \colon \R^q \to [0, \infty)$ is  
continuous and coercive.   
\end{enumerate}
\end{assumption}

All above  assumptions are naturally fulfilled in our context. 
\ref{ass:1} is  satisfied for typical network architectures, in particular  for CNNs.  \ref{ass:2} is satisfied for the patch extraction operator and, finally \ref{ass:3} is  satisfied for typically used regularizers such as the Frobenius-norm or weighted   $\ell^q$-norms.  Recall that  $\pen$ is called  
coercive if $\pen(\theta_k) \to \infty$ if $(\theta_k)_{k\in \N}$ is a sequence with 
$\norm{\theta_k}_2 \to \infty$. 

\begin{theorem}[Existence]\label{thm:ex}
Algorithm~\ref{reco_algo} defines a sequence  $(\xx_k)_{k\in \N}$ 
of iterates and a sequence of   parameters   
$(\theta_k)_{k  \in \N}$.      
\end{theorem}

\begin{IEEEproof}
In order to show that ALONE  defines sequences 
$(\xx_k)_{k\in \N}$, $(\theta_k)_{k  \in \N}$, it is sufficient to show that 
the functionals $\fun_{\theta, \yu,\lambda} $ and $\fun_{\xx, \yu,\lambda}$ both have at least one minimizer. Because of \ref{ass:0}, \ref{ass:1}, the   functional $\fun_{\theta, \yu,\lambda} $ is quadratic and nonnegative and therefore has  a minimizer. According to \ref{ass:1} the mapping $\theta \mapsto \net_\theta(\Ed(\xx))$  is continuous. Together  with \ref{ass:3}  this implies that
$ \fun_{\xx, \yu,\lambda}(\theta) = \norm{ \Ed \xx - \net_\theta(\Ed(\xx)) }_2^2 
 + \pen(\theta) $ is continuous and coercive. Standard arguments therefore imply the existence of minimizers. To see this, let 
$(\theta_k)_{k\in \N}$ be a sequence   with  
$\fun_{\xx, \yu,\lambda}(\theta_k) \to \inf_\theta \fun_{\xx, \yu,\lambda}(\theta)$. 
The coercivity of  $\fun_{\xx, \yu,\lambda}$ implies that  this sequence is bounded and therefore has at least one  convergent subsequence. The continuity  implies that the limit of the subsequence is a minimizer of $ \fun_{\xx, \yu}$.  
 \end{IEEEproof}

Note that that the minimizer of $\fun_{\theta, \yu,\lambda} $ is  unique if the matrix 
$\Hd$ has full rank. For example, this is the case of $ \Ed^\trans  \Ed$
has full rank. For the patch extraction operator, the full rank condition is satisfied, provided the patches cover the  whole image domain.  
However, the  minimizer of  $\fun_{\xx, \yu,\lambda}$ might be non-unique due to the non-convexity of $ \theta \mapsto \| \Ed_j (\xx) - \net_\theta\big(\Ed_j(\xx)\big) \|_2^2 $.   
Therefore ALONE  depends on the particular choice of the minimizers of $\fun_{\xx, \yu,\lambda}$ and, in general, does not define a unique sequence. 

\begin{definition}[Sets of  ALONE iterates]
For any  $\yu \in \C^m$ and  any iteration index $k \in \N$ 
we denote by $\Bd_k (\yu)$ the set of all iterates $(\xx_k, \theta_k) \in 
\C^N \times \R^q$ that are generated by the ALONE reconstruction Algorithm~\ref{reco_algo}, 
by choosing   arbitrary  minimizers  of $\fun_{\xx_{\ell},\yu,\lambda}$, $\fun_{\theta_{\ell},\yu,\lambda}$ for $\ell \in  \set{0, \dots, k-1}$. 
\end{definition}

We  next show that the sets of ALONE iterates $\Bd_k (\yu)$   depend stably on the inputs $\yu$. For that purpose, we write  $\Bd  \colon \C^m  \rightrightarrows  \C^N  \times \R^q$ for  a multivalued mappings where $\Bd (\yu) \subset \C^N  \times \R^q$ and use    the following  notion of stability.

\begin{definition}[Stability of multivalued mappings]
A multivalued mapping $\Bd  \colon \C^m  \rightrightarrows  \C^N  \times \R^q$   is called stable if for any $\yu \in \C^m$ and any sequence  $(\yu^\ell)_{\ell \in \N}$  converging to 
$\yu$ the following statements hold true:  
\begin{enumerate}[label=(\roman*)]
\item $(\Bd(\yu^\ell ))_{\ell  \in \N}$ has at least one accumulation point.
 
 \item All accumulation points of   $(\Bd(\yu^\ell ))_{\ell  \in \N}$ are in $\Bd (\yu)$.
\end{enumerate}   
\end{definition}

\begin{theorem}[Stability of ALONE  iterates]\label{thm:stab}
For any iteration index $k \in \N$,  the set of ALONE iterates  
$\Bd_k$  is stable.
\end{theorem}  

\begin{IEEEproof}
An inductive argument shows that it is sufficient verify 
that  $\xx  \mapsto  \argmin_\theta   \fun_{\theta, \yu,\lambda} (\xx )$ is stable. For that purpose, let   $(\yu^\ell)_{\ell \in \N}$ be a sequence converging to  $\yu \in \C^m$ and let 
$\xx^* \in \argmin \fun_{\theta, \yu,\lambda} (\xx)$,
$\xx^\ell \in \argmin \fun_{\theta, \yu^\ell,\lambda} (\xx)$. 
Then  $\fun_{\theta, \yu^\ell,\lambda} (\xx^\ell) \leq \fun_{\theta, \yu^\ell,\lambda} (\xx^*)$ and therefore   
\begin{multline*}
\fun_{\theta, \yu,\lambda} (\xx^\ell) \\
\begin{aligned}
&= 
\frac{\lambda}{2} \norm{\Ed \xx^\ell - \ZZ}_2^2 + 
\frac{1}{2} \norm{ \Au \xx - \yu}_2^2
\\
&\leq 
\frac{\lambda}{2}  \norm{\Ed \xx^\ell - \ZZ}_2^2 + 
 \norm{ \Au \xx^\ell - \yu^\ell}_2^2 
+ \norm{\yu-\yu^\ell}_2^2
\\
&\leq 
2 \fun_{\theta, \yu^\ell,\lambda} (\xx^\ell)
+ \norm{\yu-\yu^\ell}_2^2
\\
&\leq 
2 \fun_{\theta, \yu^\ell,\lambda} (\xx^*)
+ \norm{\yu-\yu^\ell}_2^2
\\
&\leq 
4 \fun_{\theta, \yu,\lambda} (\xx^*)
+ 3 \norm{\yu-\yu^\ell}_2^2 \,.
\end{aligned}  
\end{multline*}  
Because $\norm{\yu-\yu^\ell}_2 \to 0$, this and the coercivity of 
$\fun_{\theta, \yu,\lambda}$ show that $(\xx^\ell)_{\ell \in \N}$  is 
bounded. In particular, $(\xx^\ell)_{\ell \in \N}$ has at least one accumulation point. Let   
$(\xx^{\tau (\ell)})_{\ell \in \N}$ be a  subsequence converging to some $\hat \xx \in \C^N$. The continuity of the norm implies 
 \begin{multline*}
\fun_{\theta, \yu,\lambda} (\hat \xx) 
= 
\lim_{\ell \to \infty} \fun_{\theta, \yu^{\tau (\ell)},\lambda} (\xx^{\tau (\ell)})             
\\
\leq  
\liminf_{\ell \to \infty}  \fun_{\theta, \yu^{\tau (\ell)},\lambda} (\xx^*)     
=
\fun_{\theta, \yu,\lambda} (\xx^*) 
= 
\inf_\xx \fun_{\theta, \yu,\lambda} (\xx)  \,. 
\end{multline*}
Consequently, $\hat \xx \in \argmin_\xx \fun_{\theta, \yu,\lambda} (\xx)$.   
\end{IEEEproof}

Clearly a main theoretical questions is to show, under suitable assumptions,  convergence of ALONE to fixed points defined by \eqref{eq:fix1}, \eqref{eq:fix2}. This turned out to be a very challenging question that we aim to investigate  in future work.

\bibliographystyle{IEEEtran}
\bibliography{IEEEabrv,references}

\begin{thebibliography}{10}
\providecommand{\url}[1]{#1}
\csname url@samestyle\endcsname
\providecommand{\newblock}{\relax}
\providecommand{\bibinfo}[2]{#2}
\providecommand{\BIBentrySTDinterwordspacing}{\spaceskip=0pt\relax}
\providecommand{\BIBentryALTinterwordstretchfactor}{4}
\providecommand{\BIBentryALTinterwordspacing}{\spaceskip=\fontdimen2\font plus
\BIBentryALTinterwordstretchfactor\fontdimen3\font minus
  \fontdimen4\font\relax}
\providecommand{\BIBforeignlanguage}[2]{{%
\expandafter\ifx\csname l@#1\endcsname\relax
\typeout{** WARNING: IEEEtran.bst: No hyphenation pattern has been}%
\typeout{** loaded for the language `#1'. Using the pattern for}%
\typeout{** the default language instead.}%
\else
\language=\csname l@#1\endcsname
\fi
#2}}
\providecommand{\BIBdecl}{\relax}
\BIBdecl

\bibitem{weiger2000cardiac}
M.~Weiger, K.~P. Pruessmann, and P.~Boesiger, ``Cardiac real-time imaging using
  sense,'' \emph{Magnetic Resonance in Medicine: An Official Journal of the
  International Society for Magnetic Resonance in Medicine}, vol.~43, no.~2,
  pp. 177--184, 2000.

\bibitem{lin2004parallel}
F.-H. Lin, K.~K. Kwong, J.~W. Belliveau, and L.~L. Wald, ``Parallel imaging
  reconstruction using automatic regularization,'' \emph{Magnetic Resonance in
  Medicine: An Official Journal of the International Society for Magnetic
  Resonance in Medicine}, vol.~51, no.~3, pp. 559--567, 2004.

\bibitem{candes2008introduction}
E.~J. Cand{\`e}s and M.~B. Wakin, ``An introduction to compressive sampling,''
  \emph{IEEE Signal Processing magazine}, vol.~25, no.~2, pp. 21--30, 2008.

\bibitem{donoho2006compressed}
D.~L. Donoho \emph{et~al.}, ``Compressed sensing,'' \emph{IEEE Transactions on
  Information Theory}, vol.~52, no.~4, pp. 1289--1306, 2006.

\bibitem{seiberlich2011improved}
N.~Seiberlich, G.~Lee, P.~Ehses, J.~L. Duerk, R.~Gilkeson, and M.~Griswold,
  ``Improved temporal resolution in cardiac imaging using through-time spiral
  grappa,'' \emph{Magnetic Resonance in Medicine}, vol.~66, no.~6, pp.
  1682--1688, 2011.

\bibitem{winkelmann2006optimal}
S.~Winkelmann, T.~Schaeffter, T.~Koehler, H.~Eggers, and O.~Doessel, ``An
  optimal radial profile order based on the golden ratio for time-resolved
  mri,'' \emph{IEEE Transactions on Medical Imaging}, vol.~26, no.~1, pp.
  68--76, 2006.

\bibitem{block2007undersampled}
K.~T. Block, M.~Uecker, and J.~Frahm, ``Undersampled radial mri with multiple
  coils. iterative image reconstruction using a total variation constraint,''
  \emph{Magnetic Resonance in Medicine}, vol.~57, no.~6, pp. 1086--1098, 2007.

\bibitem{ravishankar2010mr}
S.~Ravishankar and Y.~Bresler, ``Mr image reconstruction from highly
  undersampled k-space data by dictionary learning,'' \emph{IEEE Transactions
  on Medical Imaging}, vol.~30, no.~5, pp. 1028--1041, 2010.

\bibitem{caballero2014dictionary}
J.~Caballero, A.~N. Price, D.~Rueckert, and J.~V. Hajnal, ``Dictionary learning
  and time sparsity for dynamic mr data reconstruction,'' \emph{IEEE
  Transactions on Medical Imaging}, vol.~33, no.~4, pp. 979--994, 2014.

\bibitem{wang2014compressed}
Y.~Wang and L.~Ying, ``Compressed sensing dynamic cardiac cine mri using
  learned spatiotemporal dictionary,'' \emph{IEEE Transactions on Biomedical
  Engineering}, vol.~61, no.~4, pp. 1109--1120, 2014.

\bibitem{jin2017deep}
K.~H. Jin, M.~T. McCann, E.~Froustey, and M.~Unser, ``{Deep Convolutional
  Neural Network for Inverse Problems in Imaging},'' \emph{IEEE Transactions on
  Image Processing}, 2017.

\bibitem{han2018framing}
Y.~Han and J.~C. Ye, ``Framing u-net via deep convolutional framelets:
  Application to sparse-view ct,'' \emph{IEEE Transactions on Medical Imaging},
  vol.~37, no.~6, pp. 1418--1429, 2018.

\bibitem{Hauptmann2019}
A.~Hauptmann, S.~Arridge, F.~Lucka, V.~Muthurangu, and J.~A. Steeden,
  ``Real-time cardiovascular mr with spatio-temporal artifact suppression using
  deep learning--proof of concept in congenital heart disease,'' \emph{Magnetic
  Resonance in Medicine}, vol.~81, no.~2, pp. 1143--1156, 2019.

\bibitem{kofler2019}
A.~{Kofler}, M.~{Dewey}, T.~{Schaeffter}, C.~{Wald}, and C.~{Kolbitsch},
  ``Spatio-temporal deep learning-based undersampling artefact reduction for 2d
  radial cine mri with limited training data,'' \emph{IEEE Transactions on
  Medical Imaging}, no. {DOI:} 10.1109/TMI.2019.2930318, 2019.

\bibitem{adler2017solving}
J.~Adler and O.~{\"O}ktem, ``Solving ill-posed inverse problems using iterative
  deep neural networks,'' \emph{Inverse Problems}, vol.~33, no.~12, p. 124007,
  2017.

\bibitem{adler2018learned}
------, ``Learned primal-dual reconstruction,'' \emph{IEEE Transactions on
  Medical Imaging}, vol.~37, no.~6, pp. 1322--1332, 2018.

\bibitem{hammernik2018learning}
K.~Hammernik, T.~Klatzer, E.~Kobler, M.~P. Recht, D.~K. Sodickson, T.~Pock, and
  F.~Knoll, ``Learning a variational network for reconstruction of accelerated
  mri data,'' \emph{Magnetic Resonance in Medicine}, vol.~79, no.~6, pp.
  3055--3071, 2018.

\bibitem{schlemper2017deep}
J.~Schlemper, J.~Caballero, J.~V. Hajnal, A.~N. Price, and D.~Rueckert, ``A
  deep cascade of convolutional neural networks for dynamic mr image
  reconstruction,'' \emph{IEEE Transactions on Medical Imaging}, vol.~37,
  no.~2, pp. 491--503, 2018.

\bibitem{qin2018convolutional}
C.~Qin, J.~Schlemper, J.~Caballero, A.~N. Price, J.~V. Hajnal, and D.~Rueckert,
  ``Convolutional recurrent neural networks for dynamic mr image
  reconstruction,'' \emph{IEEE Transactions on Medical Imaging}, vol.~38,
  no.~1, pp. 280--290, 2018.

\bibitem{Feng2015}
L.~Feng, L.~Axel, H.~Chandarana, K.~T. Block, D.~K. Sodickson, and R.~Otazo,
  ``{XD-GRASP : Golden-Angle Radial MRI with Reconstruction of Extra
  Motion-State Dimensions Using Compressed Sensing},'' \emph{Magn. Reson.
  Med.}, vol.~00, no. October 2014, pp. 1--14, 2015.

\bibitem{feng_mrm_2012}
\BIBentryALTinterwordspacing
L.~Feng, M.~B. Srichai, R.~P. Lim, A.~Harrison, W.~King, G.~Adluru, E.~V.~R.
  Dibella, D.~K. Sodickson, R.~Otazo, and D.~Kim, ``{Highly accelerated
  real-time cardiac cine MRI using k-t SPARSE-SENSE.}'' \emph{Magn. Reson.
  Imag.}, aug 2012. [Online]. Available:
  \url{http://dx.doi.org/10.1002/mrm.24440}
\BIBentrySTDinterwordspacing

\bibitem{lin2018python}
J.-M. Lin, ``Python non-uniform fast fourier transform (pynufft): An
  accelerated non-cartesian mri package on a heterogeneous platform
  (cpu/gpu),'' \emph{Journal of Imaging}, vol.~4, no.~3, p.~51, 2018.

\bibitem{kingma2014adam}
D.~P. Kingma and J.~Ba, ``Adam: A method for stochastic optimization,''
  \emph{arXiv preprint arXiv:1412.6980}, 2014.

\bibitem{hestenes1952methods}
M.~R. Hestenes, E.~Stiefel \emph{et~al.}, ``Methods of conjugate gradients for
  solving linear systems,'' \emph{Journal of research of the National Bureau of
  Standards}, vol.~49, no.~6, pp. 409--436, 1952.

\bibitem{Tsao2003}
J.~Tsao, P.~Boesiger, and K.~P. Pruessmann, ``k-t blast and k-t sense: dynamic
  mri with high frame rate exploiting spatiotemporal correlations,''
  \emph{Magnetic Resonance in Medicine: An Official Journal of the
  International Society for Magnetic Resonance in Medicine}, vol.~50, no.~5,
  pp. 1031--1042, 2003.

\bibitem{reisenhofer2018haar}
R.~Reisenhofer, S.~Bosse, G.~Kutyniok, and T.~Wiegand, ``A haar wavelet-based
  perceptual similarity index for image quality assessment,'' \emph{Signal
  Processing: Image Communication}, vol.~61, pp. 33--43, 2018.

\bibitem{chambolle2005total}
A.~Chambolle, ``Total variation minimization and a class of binary mrf
  models,'' in \emph{International Workshop on Energy Minimization Methods in
  Computer Vision and Pattern Recognition}.\hskip 1em plus 0.5em minus
  0.4em\relax Springer, 2005, pp. 136--152.

\bibitem{wang2004image}
Z.~Wang, A.~C. Bovik, H.~R. Sheikh, E.~P. Simoncelli \emph{et~al.}, ``Image
  quality assessment: from error visibility to structural similarity,''
  \emph{IEEE Transactions on Image Processing}, vol.~13, no.~4, pp. 600--612,
  2004.

\bibitem{schnass2018convergence}
K.~Schnass, ``Convergence radius and sample complexity of itkm algorithms for
  dictionary learning,'' \emph{Applied and Computational Harmonic Analysis},
  vol.~45, no.~1, pp. 22--58, 2018.

\bibitem{aharon2006k}
M.~Aharon, M.~Elad, A.~Bruckstein \emph{et~al.}, ``K-svd: An algorithm for
  designing overcomplete dictionaries for sparse representation,'' \emph{IEEE
  Transactions on Signal Processing}, vol.~54, no.~11, p. 4311, 2006.

\bibitem{tropp2007signal}
J.~A. Tropp and A.~C. Gilbert, ``Signal recovery from random measurements via
  orthogonal matching pursuit,'' \emph{IEEE Transactions on Information
  Theory}, vol.~53, no.~12, pp. 4655--4666, 2007.

\end{thebibliography}

\end{document}